\documentclass{aastex631}

\usepackage{longtable}
\usepackage{multirow}
\usepackage{amsmath}
\usepackage{xcolor}
\usepackage{subfigure}

\definecolor{green1}{rgb}{0.2, 0.6, 0.1}
\definecolor{purp1}{rgb}{0.6, 0.1, 0.45}

\def\d{{\rm d}}

\newcommand{\dL}{$\delta \mathcal{L}^{\prime}$}
\newcommand{\dLc}{$\delta \mathcal{L}_c^{\prime}$}

\newcommand{\Lc}{$\mathcal{L}_c^{\prime}$}
\newcommand{\Hc}{$\mathcal{H}_c^{\prime}$}

\newcommand{\dH}{$\delta \mathcal{H}^{\prime}$}
\newcommand{\dHc}{$\delta \mathcal{H}_c^{\prime}$}

\newcommand{\vzdl}{$v_z\,\delta \mathcal{L}_c^{\prime}$}
\newcommand{\vzdh}{$v_z\,\delta \mathcal{H}_c^{\prime}$}
\newcommand{\vzbz}{$v_z\left|B_z\right|$}

\newcommand{\kurtvzdl}{$\mathrm{Kurt}\left( v_z\,\delta \mathcal{L}_c^{\prime} \right)$}
\newcommand{\kurtvzdh}{$\mathrm{Kurt}\left( v_z\,\delta \mathcal{H}_c^{\prime} \right)$}

\def\d{\mathrm{d}}

\newcommand{\FARc}{110}                     
\newcommand{\NFc}{122}                      
\newcommand{\FAR}{57}                       
\newcommand{\NF}{175}                       
\newcommand{\TF}{232}                       
\newcommand{\cflares}{2142}                 
\newcommand{\mflares}{384}                  
\newcommand{\xflares}{36}                   

\newcommand{\ValidationNumber}{25}          
\newcommand{\TrainingNumber}{207}           
\newcommand{\HoldoutNumber}{12}              
\newcommand{\percVal}{$10\,\%$}             
\newcommand{\percTrain}{$90\,\%$}           
\newcommand{\predictionRange}{24}           
\newcommand{\historicalRange}{24 and 12}    
\newcommand{\laggedTimes}{0.2 and 0.4}      
\newcommand{\runningTimes}{2}               

\newcommand{\tsssa}{0.524}                 
\newcommand{\TSSalert}{0.521}              

\received{December 16, 2025}
\submitjournal{ApJ}

\shorttitle{XGBoost Model}
\shortauthors{Williams et al.}

\begin{document} 
    \correspondingauthor{Thomas Williams}
    \email{tomwilliamsphd@gmail.com}
    
   \title{Investigating the Efficacy of Topologically Derived Time Series for Flare Forecasting. II. XGBoost Model}

   \author[0000-0002-2006-6096]{Thomas Williams}
   \affiliation{Department of Mathematical Sciences, Durham University, Durham, UK}

   \author[0000-0003-4015-5106]{Christopher B. Prior}
   \affiliation{Department of Mathematical Sciences, Durham University, Durham, UK}

   \author[0000-0003-2297-9312]{David MacTaggart}
   \affiliation{School of Mathematics \& Statistics, University of Glasgow, Glasgow, UK}

   \author[0000-0001-9521-3874]{O.P.M. Aslam}
   \affiliation{School of Mathematics \& Statistics, University of Glasgow, Glasgow, UK}

   \author[0000-0002-4183-9895]{D. Shaun Bloomfield}
   \affiliation{School of Engineering, Physics and Mathematics, Northumbria University, Newcastle upon Tyne, UK}

\begin{abstract}
Solar flares are a primary driver of space weather, and forecasting their occurrence remains a significant challenge. This paper presents a novel flare prediction model based on topologically derived photospheric magnetic parameters. We employ the \texttt{ARTop} framework to compute the time-dependent input rates of magnetic winding and helicity across more than $10^5$ active region (AR) observations, decomposing them into current-carrying and potential components to reduce sensitivity to optical flow methods. An \texttt{XGBoost} machine learning model is trained on these topological time series, alongside engineered features including rolling statistics, kurtosis, and flare history, to predict the probability of $\geq$M1.0-class flares within the next 24 hours. The model demonstrates strong performance on a validation set, with a True Skill Statistic (TSS) of 0.804 for once daily operational region forecasts. When applied to a fully independent holdout set, the operational forecast achieves a TSS of \tsssa. A SHapley Additive exPlanations (SHAP) analysis confirms the model's physical interpretability, identifying flare history and accumulated current-carrying winding and helicity as the most important features. The main challenges identified are false positives arising from ARs with frequent C-class flaring and systematic errors introduced by projection effects when ARs are near the limb. Excluding limb-affected data yields no improvement in the holdout set TSS (\TSSalert\ versus \tsssa), due to the overall decreased number of flares. However, our per-region analysis indicates that mitigating these projection effects is crucial for future operational deployment. This work establishes magnetic topology, particularly its current-carrying components, as a highly effective and physically meaningful set of predictors for solar flare forecasting.
\end{abstract}
\keywords{solar flares -- space weather -- solar active region magnetic fields -- astronomy databases -- machine learning}

\section{Introduction}\label{sec:intro}
Solar flares are intense, impulsive releases of magnetic energy in the solar corona, characterized by sudden enhancements in electromagnetic radiation across the spectrum -- from radio waves through ultraviolet (UV) to X-rays and gamma rays. These events originate in magnetically complex active regions and produce copious non-thermal electron and ion acceleration, which drive phenomena such as solar radio bursts and EUV/X-ray enhancements \citep{benz17}. The rapid ionization of Earth's lower ionosphere (notably the D- and E-layers) produces sudden ionospheric disturbances (SIDs), often referred to as the magnetic crochet effect, which can severely degrade high-frequency (HF) radio communication and disrupt Global Navigation Satellite System (GNSS) signals \citep{Temmer2021,Gopalswamy2022a}. In addition, flare-associated enhancements in ultraviolet and soft X-ray flux increase atmospheric heating, expanding the upper atmosphere and enhancing drag on low-Earth-orbit satellites, thereby affecting satellite orbit maintenance and lifetime \citep{Gopalswamy2022a}.

In the broader landscape of space weather, solar flares frequently occur with coronal mass ejections (CMEs) and solar energetic particle (SEP) events to produce compounded effects in near-Earth space. While CMEs drive geomagnetic storms via magnetospheric interactions, flares contribute prompt injections of high-energy particles that may arrive within tens of minutes, posing acute risks to satellite electronics, astronaut radiation exposure, and commercial aviation at high latitudes. Notably, SEP events accelerate via both flare-associated reconnection and CME-driven shocks, underscoring the necessity to model both mechanisms in operational forecasting \citep{Klein2017}. Subsequently, solar flares constitute critical drivers of electromagnetic, ionospheric, and radiative disruptions in the Sun-Earth system and remain primary targets for predictive machine learning (ML) pipelines in operational space weather models.

The flux of magnetic helicity has long been considered an important quantity in the study of active regions and solar flares \citep{pevtsov2003helicity,park2008variation,vemareddy2021successive,korsos2022magnetic,liu2023changes}.Another similar quantity known as the magnetic winding which describes the geometric twisting of more horizontal magnetic field lines is a relativity novel quantity that has also been shown to have additional predictive efficacy \citep{prior2020magnetic,mactaggart21,brenopaper,mnraspaper}.

\citet[hereafter WPM]{williams25} leverage the Active Region Topology (\texttt{ARTop}) framework to generate time-resolved metrics of magnetic helicity flux ${\cal H}'$ and winding input ${\cal L}'$ rates derived from Helioseismic and Magnetic Imager \citep[HMI]{hmipaper} data aboard the Solar Dynamics Observatory (SDO) in the form of Space-Weather HMI Active Region Patches (SHARP). The aim of this study is to assess the predictive potential of topological quantities with respect to solar flares. WPM focuses on imbalance measures -- specifically the absolute difference in current-carrying versus potential components of topological inputs (\dL\ and \dH) -- which were shown to be comparatively insensitive to velocity-smoothing choices and correlate strongly with flaring activity. Through analysis of a comprehensive dataset of 144 active regions (including both flaring and non-flaring cases), the study revealed that spikes in \dL\ and \dH\ time series -- including combinations with line-of-sight velocity (\vzdl\ and \vzdh) -- often precede M- and X-class flares by up to several hours, with statistically significant deviations beyond 2$\sigma$ envelopes in their running mean. These findings suggested that topologically derived signals, especially those representing current-carrying flux imbalance, offer promising early indicators of major solar eruptive events and warrant further investigation in solar flare forecasting models.

In recent years, numerous studies have adopted a ML approach in attempts to predict solar flares \citep{nishizuka2017,nishizuka21,sinha22,yi22,yun24,li25}. For example, \citet{bobra2014} utilise 4 years' of HMI SHARP data to train a Support Vector Machine (SVM) classifier that is trained on 25 derived features of active regions from which they conclude only a few of these parameters are needed to predict flaring. Similarly, \citet{florios18} compare SVM, Random Forest (RF) and multi-layer perceptrons (MLP) methods across 5 years of solar cycle 24. Their findings are that RF performs best, and when flares exceeding M1.0 are considered, they obtain a True Skill Statistic (TSS) score of 0.74 provided a probability threshold of 0.15 is specified for a dataset akin to a holdout set. Whilst other methods adopt convolutional neural networks \citep[CNN]{sande25}, recurrent neural networks \citep[RNN]{pandey23} or some combination of the two \citep{guastavino2022} for the basis of their predictive models, these studies incorporate advanced ML frameworks explicitly targeting forecasting capability instead of mere pattern recognition and align their findings with operational space weather goals such as 6 -- 72\,hr flare prediction and full-disk or near-real-time applicability on real solar data. Though as is common in all flare prediction studies, they suffer from highly imbalanced data that can impact model training and skew skill scores and model evaluation.
    
Another algorithm rooted in advanced ML frameworks in the Extreme Gradient Boosting (\texttt{XGBoost}) method is a high-performance, scalable implementation of gradient boosting decision trees that has gained widespread adoption across various ML tasks, including time-series forecasting. It builds an ensemble of decision trees in a sequential manner that are optimised and support both first- and second-order gradient information, including regularisation to control model complexity \citep{chen2016xgboost}. For time-series and classification tasks -- such as forecasting solar flares based on SHARP parameters -- \texttt{XGBoost} has demonstrated robust performance by effectively handling nonlinear relationships, missing data, and feature importance interpretation \citep{cinto20}. Similarly, a recent comparative study by \citet{bringewald25} found that \texttt{XGBoost} performed strongly compared to RF and k-Nearest Neighbour models when classifying solar flares into GOES categories (B, C, M, X), and that \texttt{XGBoost} benefits significantly from increased dimensionality with ``state-of-the-art'' performance across numerous data science tasks \citep{chen2016xgboost}.

For the purpose of forecasting solar flares we will define explicitly what we constitute as a prediction. Historically, predictive models have utilised a broad range of time-frames between $6-48$\,hr (for example, Table\,2 in \citealp{bobra2014}). However, operationally, these forecasts or predictions on the likelihood of flaring are made once per day and give the outcome for the following 24\,hr. As such, more recent predictive methods now align with these constraints \citep[for example]{sinha22,pandey23,deshmukh23}. It is with these operational capabilities in mind that we will also adopt a 24\,hr cadence (at midnight) for making predictions of flaring for the upcoming day.

This second paper focuses upon the creation of a supervised ML model with \texttt{XGBoost} to predict the likelihood of M- and X-class flaring within the next 24\,hours using the dataset outlined in WPM. In \S\,\ref{sec:Dataset}, the \texttt{ARTop} code and living dataset, and data preparation employed for this study are outlined. \S\,\ref{sec:XGB} discusses the \texttt{XGBoost} model, the feature engineering, hyperparameter and model tuning, and the model evaluation metrics, whilst the model is evaluated on validation data in \S\,\ref{sec:ModelEval}. Our main results and analysis are presented in \S\,\ref{sec:results} and concluding remarks are provided in \S\,\ref{sec:conc}.

\section{Dataset}\label{sec:Dataset}

\subsection{Active Region Topology}\label{sec:artop}
\begin{figure}[h!]
\centering
\subfigure[]{\includegraphics[width=5cm]{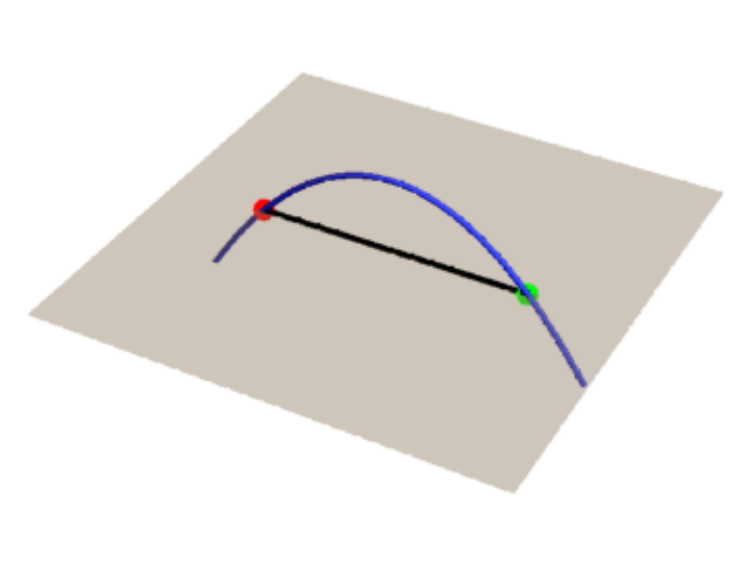}}\quad
\subfigure[]{\includegraphics[width=5cm]{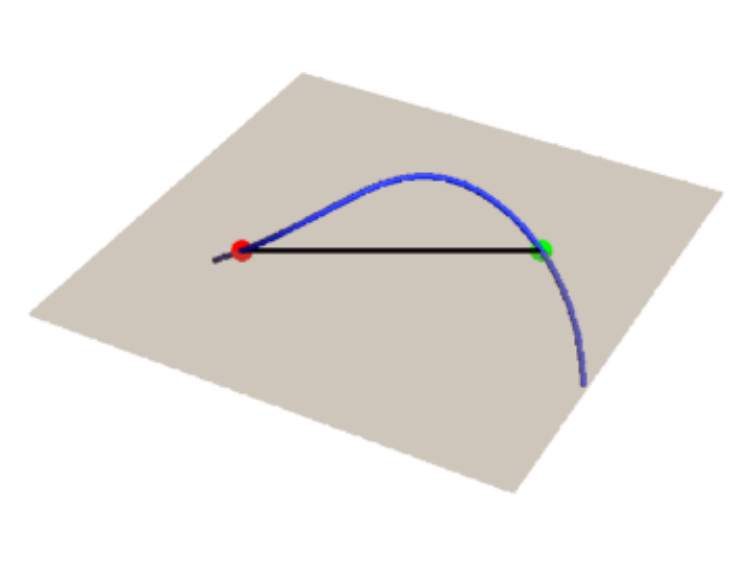}}\quad
\subfigure[]{\includegraphics[width=5cm]{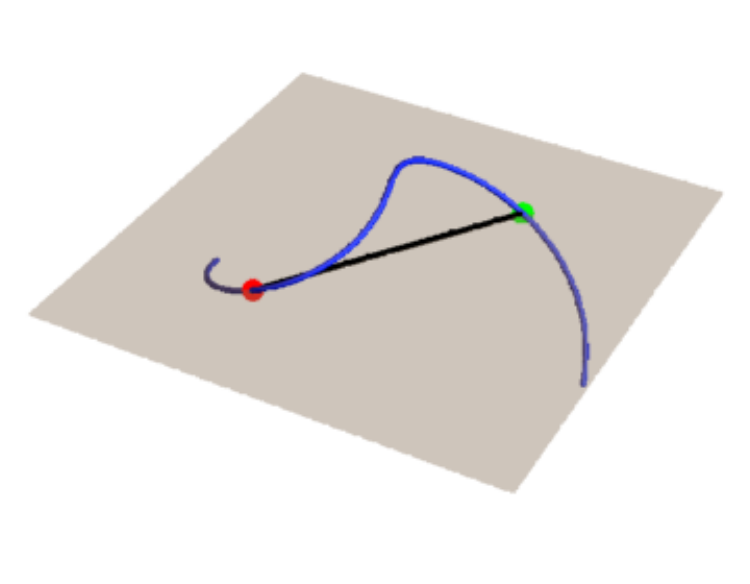}}\\[1em]
\subfigure[]{\includegraphics[width=5cm]{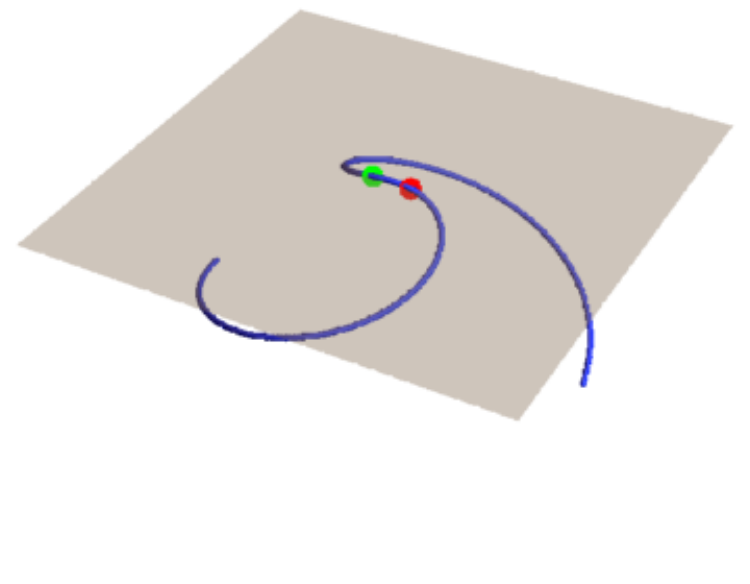}}\quad
\subfigure[]{\includegraphics[width=5cm]{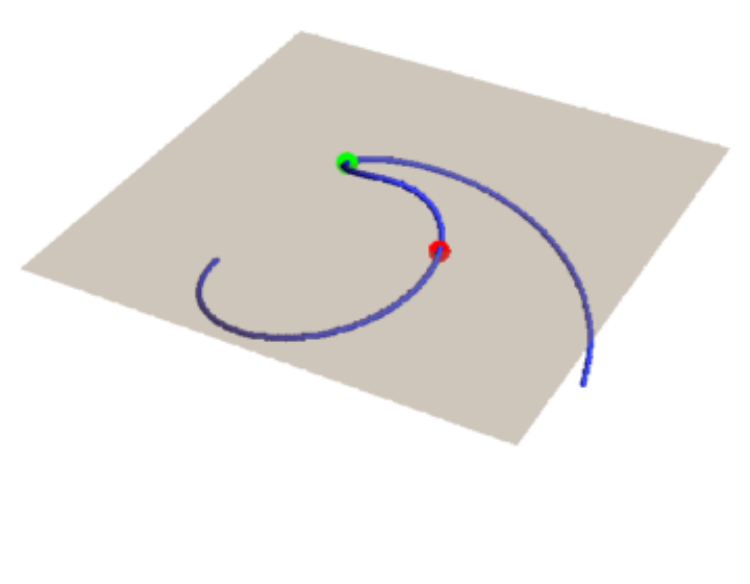}}\quad
\subfigure[]{\includegraphics[width=5cm]{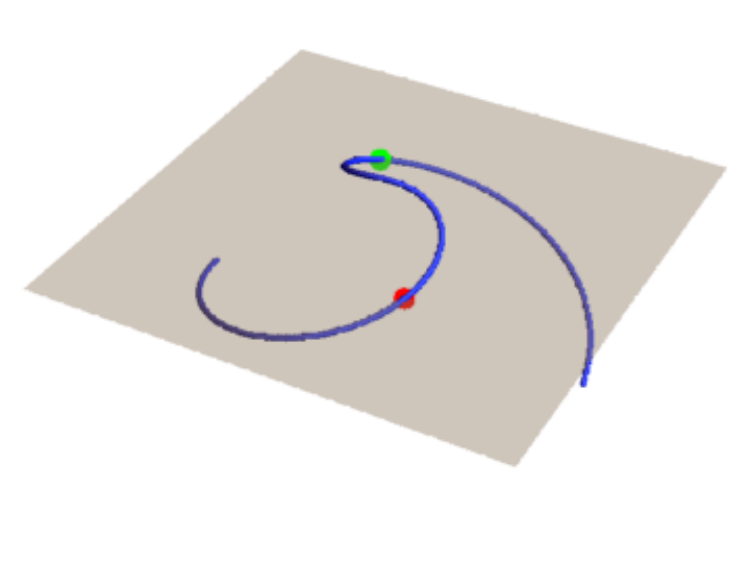}}
\caption{\label{fig:windfig}Illustrations of field properties measured by the winding flux ${\cal L}'$ and helicity flux  ${\cal H}'$. Panels (a)-(c)  depict the entanglement of a field line due to a rotational fluid motion at the photosphere (depicted as a plane). The winding measures the rotation of the green and red points where the field line pierces the plane. Panels (d)-(f) depict the rotational motion evaluated by the winding due to a contorted filed line emerging through the photosphere. The helicity measures these rotations multiplied by the flux in the field lines.}
\end{figure}
Active Region Topology \citep[\texttt{ARTop}]{artop1} is an open-source tool for studying the input of topological quantities into solar active regions at the photospheric level. \texttt{ARTop} utilises vector magnetograms \citep{hoeksema14} from HMI SHARP data to create maps, time series, and other metrics derived from input rates of magnetic helicity and magnetic winding fluxes (see \citealp{prior2020magnetic} for a detailed summary of their meaning and importance in solar applications). Figure \ref{fig:windfig} illustrates the geometrical and physical interpretation of these fluxes. Motions of field lines at the solar surface are estimated from the magnetogram data and the DAVE4VM code for the plasma velocity. This is used to estimate the mutual rotation of individual and mutual pairs of field lines at the solar surface, due to either in-plane plasma motion braiding the field, or the apparent field line motion due to field lines with complex topology emerging from the solar interior/submerging from the corona. The winding flux distribution ${\cal L}'$ at a given point in the photosphere is the average rotational motion of \textit{all} field lines around that point. The helicity flux is the winding flux weighted by the field's $B_z$ flux (field strength weighted winding). The quantities \dL\ and \dH\ are calculated from these distributions as described above. The crucial difference between the winding and helicity related quantities is that the winding quantities tend to be dominated by field around the polarity inversion line(s) whilst the helicity is dominated more by the field at the major footpoints (sunspots). For more comprehensive definitions of these quantities and the practical matter of their calculation in \texttt{ARTop}, please refer to WPM, \citet{artop1}, and references therein. 

One of the main results from WPM is that the decomposition of magnetic winding (and magnetic helicity) into components for current-carrying and potential topology are sensitive to the velocity smoothing employed in the \textit{DAVE4VM} code. Given the lack of sensitivity of \dL\ to these parameters, we instead split \dL\ and \dH\ into positive and negative components whereby positive (negative) would indicate that the dominant input of topology at the photospheric level is current-carrying (potential) topology. This is achieved by the following conditional expressions:
\begin{equation}\label{eq:dLc_dHc}
   \delta\mathcal{L}_c^{\prime} =
    \begin{cases}
       \delta\mathcal{L}^{\prime}, & \text{if } \delta\mathcal{L}^{\prime} > 0 \\
        0,                                       & \text{otherwise}
\end{cases}
\quad\text{,}\quad
    \delta\mathcal{H}_c^{\prime} =
    \begin{cases}
       \delta\mathcal{H}^{\prime}, & \text{if } \delta\mathcal{H}^{\prime} > 0 \\
        0,                                       & \text{otherwise}
\end{cases}
,
\end{equation}
and
\begin{equation}\label{eq:dLp_dHp}
    \delta\mathcal{L}_p^{\prime} =
    \begin{cases}
        \delta\mathcal{L}^{\prime}, & \text{if } \delta\mathcal{L}^{\prime} < 0 \\
        0,                                       & \text{otherwise}
\end{cases}
\quad\text{,}\quad
    \delta\mathcal{H}_p^{\prime} =
    \begin{cases}
        \delta\mathcal{H}^{\prime}, & \text{if } \delta\mathcal{H}^{\prime} < 0 \\
        0,                                       & \text{otherwise}
\end{cases}
.
\end{equation}
This subsequently reduces the sensitivity of the velocity smoothing profile selected, whilst still allowing for current-carrying decompositions to be investigated. These current-carrying components of the helicity and winding may then be incorporated alongside the other topological quantities of our dataset.

\subsection{Living Dataset}\label{sec:data}
The \TF\ SHARP regions utilised in this study are curated from the dataset outlined in WPM and \citet{hollanda21}. We ensured a number of X-class flaring regions are included in this curation as they are relatively rare amongst the dataset as a whole, but form an important target for predictive purposes. Other regions included in this study have been sampled at intervals such as to allow an even coverage across the solar cycle. As the dataset developed for this manuscript is a living dataset, it will continuously evolve as more of the regions outlined by \citet{hollanda21} are processed with \texttt{ARTop}. The reason \TF\ SHARP regions are selected for this study is the result of the computing resources required during curation with \texttt{ARTop}. As is outlined in WPM, each active region is sampled with a cadence of 720\,seconds for the entire duration it is visible within the HMI data. This cadence is needed as \texttt{ARTop} requires the full time series in order to obtain accurate integrated quantities. Additionally, \texttt{ARTop} must create 2D vector maps for velocity before it may then determine the topological quantities for each pixel of the SHARP. These calculations are expensive, with the computational costs growing exponentially for larger regions. For example, large regions like SHARP 5298 that travel across the width of the solar disk, can take over a week to process on a dedicated 32-core machine. As such, time constraints dictate our sample size for this study, as was the case with WPM where the total number of SHARP regions was 144.

Of the \TF\ SHARP regions analysed here, \FARc\ regions are considered to be flaring (C-class and above) with a further \NFc\ non-flaring regions (no flare, A and B-class). In all, the dataset presented in this manuscript contains \cflares\ X-ray flares of magnitude C1.0 and above. However, in this work, a region is only considered to be flaring if flares of X-ray class magnitude C1.0 and above are recorded in the Heliophysics Event Knowledgebase \citep[HEK]{hekpaper}. As such, this means we have \FAR\ and \NF\ flaring and non-flaring active regions with a total of \mflares\ M- and X-class flares. A list of the SHARP regions that form the validation and holdout sets are presented in Tables\,\ref{tab:validationSet} and \ref{tab:holdoutSet}. These contain \ValidationNumber\ and \HoldoutNumber\ regions with their respective strongest flare magnitudes and number of C-, M-, and X-class flares. The same information is provided in the Appendix for the training set, which is formed of \TrainingNumber\ regions.

\begin{table}
\caption{SHARP regions that form the validation set for the \texttt{XGBoost} model.}
\label{tab:validationSet}                  
\centering                          
\begin{tabular}{c c c c c c}       
\hline\hline                        
NOAA Active Region & SHARP Number & Strongest Flare & \# X--class Flares & \# M--class Flares & \# C--class Flares \\ 
\hline                             
11070 & 14 & --- & --- & --- & --- \\
11132 & 285 & --- & --- & --- & --- \\
11158 & 377 & X2.2 & 1 & 5 & 54 \\
11211 & 590 & --- & --- & --- & --- \\
11214/17 & 602 & --- & --- & --- & --- \\
11291 & 851 & --- & --- & --- & --- \\
11429/30 & 1449 & X5.4 & 3 & 14 & 36 \\
11437 & 1480 & --- & --- & --- & --- \\
11469/73 & 1611 & C2.6 & --- & --- & 12 \\
11488 & 1688 & --- & --- & --- & --- \\
11546 & 1942 & --- & --- & --- & --- \\
11561 & 1997 & --- & --- & --- & --- \\
11572 & 2036 & --- & --- & --- & --- \\
11640 & 2337 & C4.0 & --- & --- & 6 \\
11829 & 3113 & C1.6 & --- & --- & 1 \\
11839 & 3171 & --- & --- & --- & --- \\
11893/900 & 3364 & X1.0 & 1 & 4 & 33 \\
11910 & 3441 & --- & --- & --- & --- \\
11922 & 3490 & --- & --- & --- & --- \\
12003/06 & 3845 & C7.3 & --- & --- & 10 \\
12111 & 4328 & --- & --- & --- & --- \\
12297 & 5298 & X2.1 & 1 & 22 & 92 \\
12661 & 7034 & C8.0 & --- & --- & 7 \\
12673 & 7115 & X9.3 & 4 & 26 & 52 \\
12683 & 7148 & --- & --- & --- & --- \\
\hline                                   
\end{tabular}
\end{table}
\begin{table}
\caption{SHARP regions that form the holdout set for the \texttt{XGBoost} model.}
\label{tab:holdoutSet}                  
\centering                          
\begin{tabular}{c c c c c c}       
\hline\hline                        
NOAA Active Region & SHARP Number & Strongest Flare & \# X--class Flares & \# M--class Flares & \# C--class Flares \\ 
\hline
11976/80 & 3730 & M3.0 & --- & 1 & 9 \\
12047/51 & 4071 & M1.8 & --- & 2 & 23 \\
12228 & 4896 & --- & --- & --- & --- \\
12282 & 5186 & M2.4 & --- & 1 & 6 \\
12322 & 5446 & M4.0 & --- & 7 & 6 \\
12449/50 & 6078 & M3.9 & --- & 1 & 6 \\
12740 & 7357 & C9.9 & --- & --- & 9 \\
13473 & 10307 & C2.4 & --- & --- & 1 \\
13841 & 11968 & M1.5 & --- & 4 & 8 \\
13920 & 12392 & M2.7 & --- & 1 & 21 \\
13922 & 12404 & M2.2 & --- & 8 & 15 \\
13924 & 12405 & M2.1 & --- & 2 & 11 \\
\hline                                   
\end{tabular}
\end{table}

\subsection{Data Preparation}
The data used in this study are derived from the living dataset outlined in \S\,\ref{sec:data}. All SHARP active regions are loaded into memory, cleaned, and separated into training and validation subsets. Importantly, a stratified sampling scheme is applied so that the validation set contains the three benchmark major flaring regions (NOAA ARs 11158, 11429, and 12673) from WPM. This ensures that the validation set is both statistically independent and astrophysically representative of high-impact events. The full list can be seen in Table\,\ref{tab:validationSet}.

In an attempt to reduce the impact of spurious measurements associated with projection effects at the limb, all snapshots where the heliographic position of the SHARP region exceeds a defined longitudinal threshold ($\pm\,60\,^{\circ}$) are flagged based on the percentage of pixels affected. The idea being that this may allow a ML model to identify trends in the spurious values in the \texttt{ARTop} calculations during training, which we will discuss later in \S\,\ref{sec:results}. Additionally, rows containing missing or corrupted values are removed, and the remaining data for each region are concatenated into unified training and validation \texttt{DataFrames}, which are subsequently stored in serialised pickle format for reproducibility and efficient downstream use. At this stage, the percentage of NaN entries and limb-contaminated samples are reported to quantify data quality (typically $<6\%$ and $<25\%$ of total samples, respectively).

\begin{figure*}
\centerline{\includegraphics[width=0.7\textwidth]{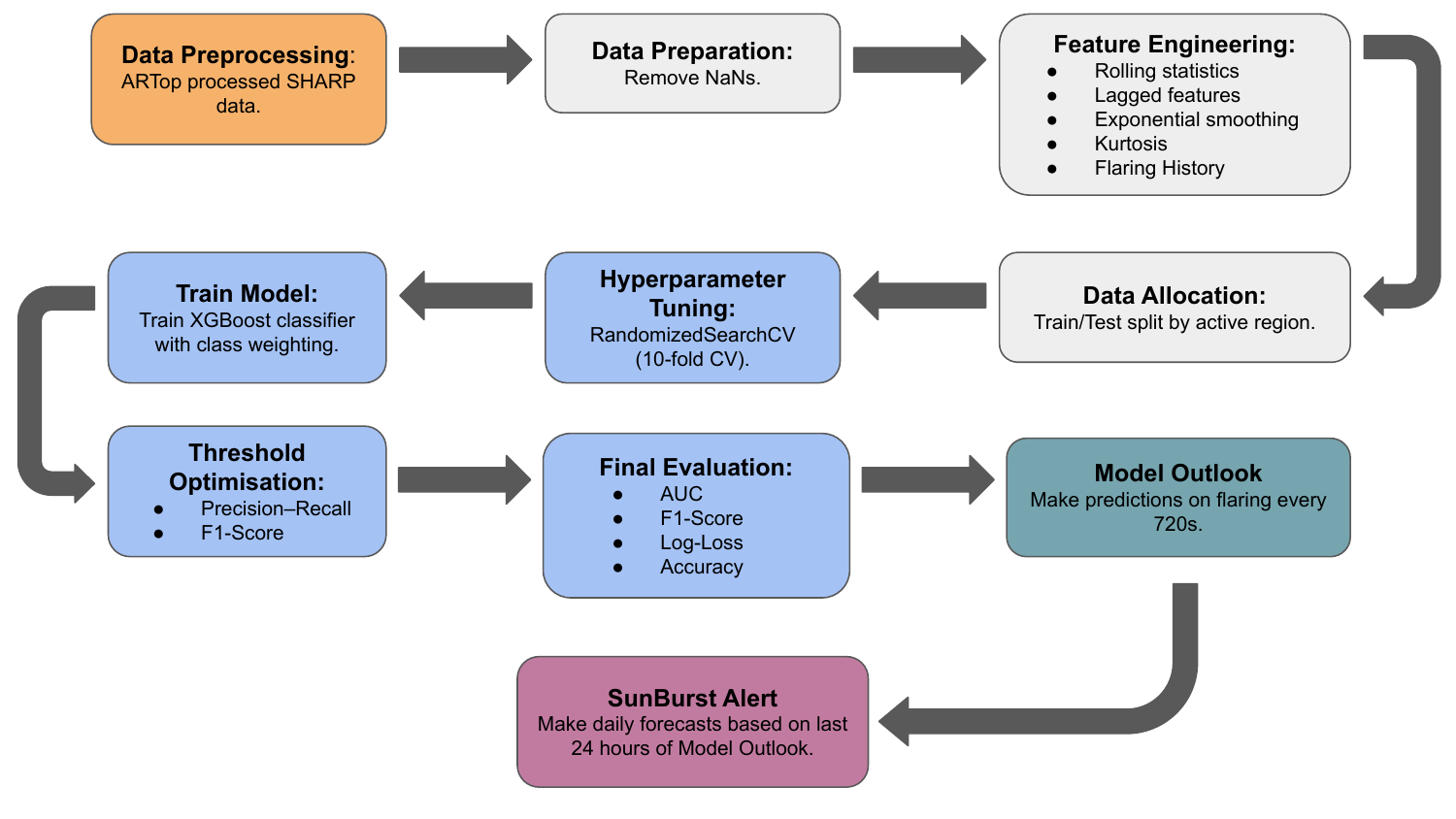}}
\caption{Workflow diagram of the \texttt{XGBoost} classification model pipeline.}
\label{fig:workflow}
\end{figure*}

\section{XGBoost Model}\label{sec:XGB}
This section outlines the supervised ML pipeline designed to forecast solar flares within a 24-hour prediction window. The pipeline integrates a series of custom data-preparation routines, feature engineering procedures, and classification using the Extreme Gradient Boosting (\texttt{XGBoost}) algorithm \citep{chen2016xgboost}. The modular framework is composed of (1) data extraction and organization, (2) feature engineering from SHARP vector magnetogram time series, (3) dataset partitioning into training and validation subsets, (4) model training with hyperparameter optimisation, and (5) evaluation using multiple performance metrics with optimised classification thresholds. The overall workflow of the model is illustrated in Figure~\ref{fig:workflow}.

\subsection{Feature Engineering}\label{sec:preproc}
A central component of the pipeline is the construction of features designed to capture both the instantaneous state and temporal evolution of the active region magnetic field by creating rolling statistics, lagged variables, and higher-order temporal descriptors. For each SHARP-derived quantity, rolling means and standard deviations are computed, capturing short-term variability. These parameters provide the model with memory of past activity.

As outlined by WPM, the quantities given by condition \ref{eq:dLc_dHc} (particularly \dLc) provide promising correlations with flares from 65 flaring active regions. As the living dataset does not contain these quantities, they are created in place of the other winding and helicity rates. Similarly in WPM, a build-up of accumulated winding and helicity seems to indicate a greater propensity for larger flares to occur within an active region. As these accumulations are based upon \dL\ and \dH, when potential-field topology is dominant for periods, this can reduce the accumulated quantities and make it more difficult to estimate the build-up of complex current-carrying topology. For this reason, the accumulated helicity and winding are also converted into the current-carrying components only. As such, the quantities of the living dataset that are incorporated into our dataset for ML are: \dLc\ and \dHc\ (and their accumulated values), \dL, \dH, \vzdl, \vzdh, and \vzbz. Additional lagged data indicating the previous behaviour at offsets of \laggedTimes\ hours alongside rolling statistics for the mean and standard deviation taken over the last \runningTimes\ hours for each quantity are also computed. 

To further smooth the high-cadence time series, single exponential smoothing \citep[SES]{holt57,brown59,winters60} is applied. These operations suppress stochastic noise while preserving physically meaningful trends. The advantage of SES compared to a standard moving mean when removing high-frequency noise is that all previous points are not weighted equally but by utilising exponentially decaying weighting factors such that more recent time series values are weighted more heavily. This weighting allows for more recent values to be deemed more important than those seen at the start of the smoothing window, giving a recency bias.

In addition to smoothing, higher-order statistics are extracted. For example, rapid increases in the excess kurtosis have been shown to be an important precursor to system bifurcations (or critical transitions) in a number of fields such as in ecological systems \citep{dakos2019ecosystem}, climate \citep{boers2018early}, the economy \citep{sevim2014developing}, and medicine \citep{chen2012detecting}. As such, kurtosis values are determined for key magnetic quantities (e.g., \dLc\ and \dHc) over 3, 6, and 12-hour windows, providing a measure of intermittency and `burstiness' in the magnetic field evolution. These descriptors were shown in WPM to correlate with flare productivity.

Furthermore, we engineer flare history features: backward-looking flare indices are calculated for the 12 and 24 hours preceding a given observation. These features quantify an active region’s recent flaring history, which is often predictive of future activity. In order to provide a proxy of this historical energy expenditure of flares within an active region, time series which quantify a score to the flaring are created. The idea being that over a period of several hours, an active region could release similar quantities of built-up magnetic energy by numerous smaller C-- and M--class flares as it might with a single X--class flare. To assign a numerical value to a flare ($f_{val}$), we follow a logarithmic classification function such that
\begin{equation}
f_{val} = 
    \begin{cases}
        10x, & \text{if C--class}\\
        100x, & \text{if M--class}\\
        1000x, & \text{if X--class}\\
        0, & \text{Otherwise}
    \end{cases}
    ,
\end{equation}
where $x$ is the flare classification's numerical suffix. For example, a C2.6 flare would be assigned a value of 26 whilst an X9.3 flare would be assigned a value of 9300. The motivation for taking a multiplier of 10 on the C-class flares allows for greater separation between the flaring and non-flaring outcomes, which may aid model training. Based upon this, the flare score time series are then calculated on running-sums of all the $f_{val}$ values utilising kernels of the previous \historicalRange\ hours, respectively. This is based on a common flare score metric \citep{antalova1996daily,abramenko2005relationship,park2010productivity}. Note that here, the ranges do not include the last hour to approximate real-time observational delays between a flare being detected and its magnitude being quantified and publicised.

To stabilize feature distributions, all variables with the exception of integrated quantities, flare labels, limb-affected pixel percentage, and kurtoses are log-transformed. Negative values are preserved by mapping their magnitudes through a signed $\log_{10}$ transform (eq.\,\ref{eq:logNorm}), ensuring the statistical properties of positive and negative excursions are comparably scaled. We chose this transformation as it reduces skew and prevents dominance of a small subset of parameters with large raw magnitudes. Mathematically, the expression is as follows:

\begin{equation}\label{eq:logNorm}
    X =
    \begin{cases}
        -\log_{10}{|X|}, & \text{if } X < 0 \\
        \log_{10}{X},    & \text{if } X > 0 \\
        0,          & \text{otherwise}
    \end{cases}
,
\end{equation}
where $X$ is the metric to normalise.

Finally, forward-looking flare labels are constructed at 24-hour horizons, defining the binary classification target variable.

\subsection{Dataset Partitioning}
Following feature engineering, the living dataset is organised into training and validation sets that contain \percTrain\ and \percVal\ of the SHARP regions, respectively. It is also worth noting that each set retained the full time series for their assigned SHARP regions to prevent temporal leakage, ensuring that validation and holdout sets purely consist of unseen data. The regions included in the validation set are outlined in Table\,\ref{tab:validationSet} whilst the training set is included in the Appendix. An additional holdout set is also included (Table\,\ref{tab:holdoutSet}) that consists of \HoldoutNumber\ active regions that are not part of the \TF\ SHARP regions which form the training and validation sets. Within each set, flare magnitudes are binarised: samples with no flare or C-class flare activity are labelled as $0$ (non-flaring), while those associated with M- or X-class flares within the next 24 hours are labelled as $1$ (flaring). This decision reflects operational needs to predict impactful space weather events, where false negatives for strong flares are particularly costly.

\subsection{Evaluation Metrics}
The final model is assessed on the validation set using five metrics. Letting TN, TP, FN, and FP denote true negative, true positive, false negative, and false positive outcomes, respectively, the metrics are:

\begin{enumerate}
    \item F1-score, defined as the harmonic mean of precision and recall:
    \begin{equation}
        \mathrm{F1} = 2 \cdot \frac{\mathrm{Precision} \cdot \mathrm{Recall}}{\mathrm{Precision} + \mathrm{Recall}},
    \end{equation}
    with
    \[
        \mathrm{Precision} = \frac{\mathrm{TP}}{\mathrm{TP} + \mathrm{FP}}, \quad
        \mathrm{Recall} = \frac{\mathrm{TP}}{\mathrm{TP} + \mathrm{FN}}.
    \]

    \item AUC (Area Under the ROC Curve), computed as
    \begin{equation}
        \mathrm{AUC} = \int_{0}^{1} \mathrm{TPR}(\mathrm{FPR}) \, d(\mathrm{FPR}),
    \end{equation}
    where $\mathrm{TPR} = \frac{\mathrm{TP}}{\mathrm{TP} + \mathrm{FN}}$ and $\mathrm{FPR} = \frac{\mathrm{FP}}{\mathrm{FP} + \mathrm{TN}}$. It is worth noting that TPR and Recall are the same metric but are named differently in the literature when referenced in F1 scores compared to other metrics.

    \item Log-loss, a measure of probability calibration:
    \begin{equation}
        \mathrm{LogLoss} = - \frac{1}{N} \sum_{i=1}^{N} \left[ y_i \log(p_i) + (1-y_i) \log(1-p_i) \right],
    \end{equation}
    where $y_i \in \{0,1\}$ are true labels and $p_i$ are predicted probabilities.

    \item Accuracy, the overall proportion of correctly classified instances:
    \begin{equation}
        \mathrm{Accuracy} = \frac{\mathrm{TP} + \mathrm{TN}}{\mathrm{TP} + \mathrm{TN} + \mathrm{FP} + \mathrm{FN}}.
    \end{equation}

    \item TSS (True Skill Statistic), also known as the Hansen-Kuiper skill score, which measures the ability to discriminate between events and non-events while accounting for imbalanced datasets:
    \begin{equation}
        \mathrm{TSS} = \mathrm{TPR} - \mathrm{FPR} = \frac{\mathrm{TP}}{\mathrm{TP} + \mathrm{FN}} - \frac{\mathrm{FP}}{\mathrm{FP} + \mathrm{TN}},
    \end{equation}
where TSS ranges from -1 to +1, with +1 indicating perfect prediction and 0 indicating no skill.
\end{enumerate}

These metrics are chosen to provide complementary perspectives: the F1-score highlights skill at rare-event detection, AUC measures discriminative ability across thresholds, log-loss evaluates probabilistic calibration, accuracy captures overall classification success, and TSS provides a threshold-independent measure of forecast skill that is particularly valuable for imbalanced datasets common in flare forecasting. Reporting this suite of metrics ensures comparability with other recent flare forecasting studies (e.g., \citealp{nishizuka2017, cinto20}).

\subsection{Model Training and Optimisation}\label{sec:ModelTuning}
We utilize the \texttt{XGBoost} classifier, a decision-tree ensemble method that constructs boosted trees in sequence, each correcting the errors of the prior ensemble. The algorithm is well-suited for structured tabular data and provides state-of-the-art performance in many classification tasks \citep{chen2016xgboost}. The classifier is configured with a logistic objective function (\texttt{binary:logistic}) and AUC and log-loss evaluation metrics. As flaring intervals only constitute a small minority of the dataset ($\approx8\,\%$), we apply a positive class weight factor (\texttt{scale\_pos\_weight}) to account for the data imbalance. This penalises misclassification of flares relative to non-flaring intervals, mitigating bias toward the majority class, i.e. the non-flaring intervals.

Hyperparameter tuning is conducted using the \texttt{RandomizedSearchCV} procedure from \texttt{scikit-learn}. Unlike \texttt{GridSearchCV}, which exhaustively explores all combinations, randomised search samples from predefined distributions, enabling memory efficient exploration of high-dimensional parameter spaces. We define distributions for the number of estimators (\texttt{n\_estimators}: 300--600), tree depth (\texttt{max\_depth}: 6--10), learning rate (\texttt{learning\_rate}: uniform in [0.05,0.25]), subsampling fraction (\texttt{subsample}: uniform in [0.7,1.0]), and minimum child weight (\texttt{min\_child\_weight}: 1--10). A total of 500 configurations are sampled for each fold of the search. From this, a stratified 10-fold cross-validation scheme is employed to ensure balanced representation of flaring and non-flaring samples across folds. Model performance is evaluated using three complementary metrics: area under the ROC curve (AUC), log-loss, and F1-score. To select the optimal hyperparameter set, we introduce a composite refit criterion. Each metric is normalised to the range [0,1], and a weighted average computed by:
\begin{equation}\label{eq:weightedCriterion}
S = 0.4 \cdot \widehat{\mathrm{AUC}} + 0.2 \cdot \widehat{\mathrm{LogLoss}} + 0.4 \cdot \widehat{\mathrm{F1}},
\end{equation}
where hats denote normalised scores. This composite score balances ranking ability, probability calibration, and rare-event sensitivity. It is worth noting here that the weighting adopted for metrics that comprise $S$ were determined through trial and error.

As mentioned earlier, the target variable is set using binary logic for whether there is flaring exceeding a GOES magnitude of M1.0 within the next \predictionRange\ hours. As \texttt{XGBoost} produces probabilistic outputs, a decision threshold must be applied to classify events. Rather than adopting the default threshold of $0.5$, we optimise the threshold on the validation set by computing precision--recall curves and select the value that maximises the F1-score. This ensures that the final classification strikes a balance between minimising false alarms (precision) and maximising flare detection (recall). The optimal threshold of 0.476 is stored alongside the trained model for operational use, and has a value closer to the default 0.5 such as used in \citet{cinto20} and \citet{nishizuka21} (0.5 and 0.4) than is deployed in other models such as \citet{florios18} (0.15).

\subsection{Operational Deployment}\label{sec:operation}
For each observation of a SHARP region, the \texttt{XGBoost} model is run and makes a prediction on the likelihood of a $\mathrm{M}1.0+$ flare occurring in the \predictionRange\,hours following that SHARP observation. For the purpose of clarity, we will henceforth refer to these instantaneous predictions as the Model Outlook. The reason for this clarification is because we also make forecasts once per UT day about whether flaring will occur the following UT day. This cadence is selected so as to be inline with how operational forecasts on space weather are made. These daily forecasts are the final outcome from the pipeline discussed here, and are subsequently referenced to as 24\,Hr operational forecasts.

In order to define an operational forecast, at midnight, our pipeline takes all the Model Outlooks from the previous 24\,hours, and if one or more of these outlooks indicate flaring then the operational forecast will be that flaring will occur on this day\footnote{It is possible that a stricter criterion may perform better, however, this would require further thresholding and optimisation whilst the intention here is to provide a proof-of-concept on how \texttt{ARTop} data can be utilised with ML. As such, a more in-depth definition for this alert will follow in another publication.}. The Model Outlook and operational forecasts are then discussed in \S\,\ref{sec:ModelEval} and \S\,\ref{sec:results} for the validation data. For the holdout set, the model is retrained using all of the training and validation sets with all the hyperparameter and threshold values determined during the initial training phase. This fully trained model is then evaluated in \S\,\ref{sec:hold}.

\section{Model Evaluation}\label{sec:ModelEval}
\begin{figure*}
\centerline{\includegraphics[width=0.5\textwidth]{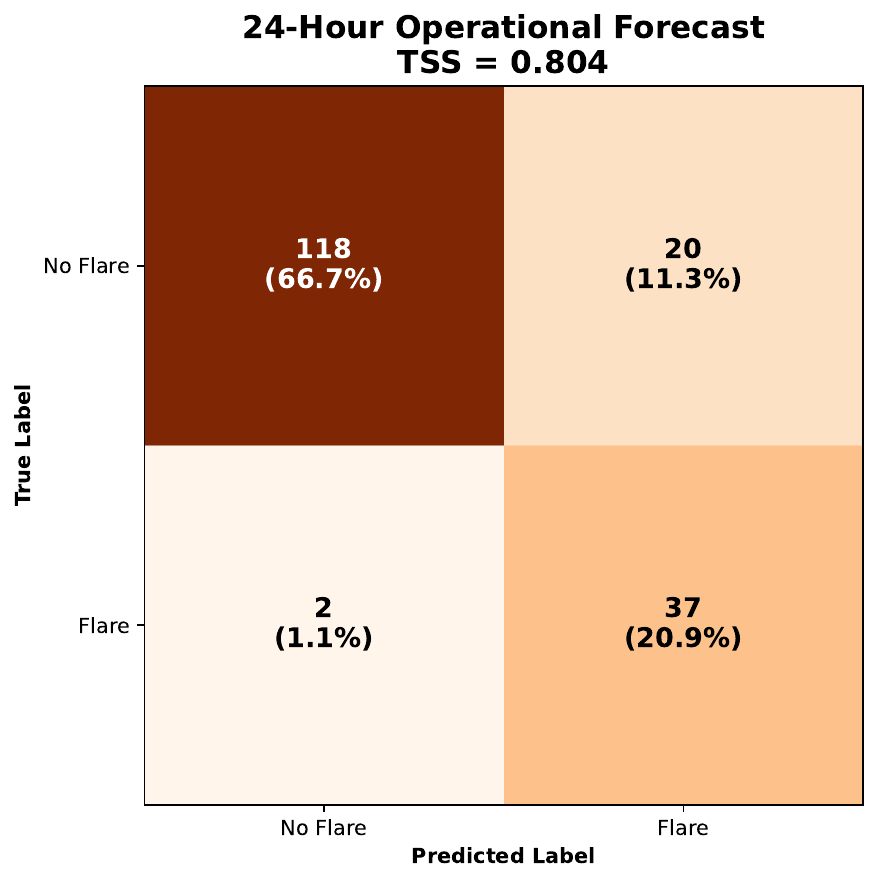}}
\caption{Confusion matrix for the \texttt{XGBoost} operational forecasts for all SHARP data in the validation set. The True Skill Statistic score is shown for illustration.}
\label{fig:confusion}
\end{figure*}
\begin{figure*}
\centerline{\includegraphics[width=\textwidth]{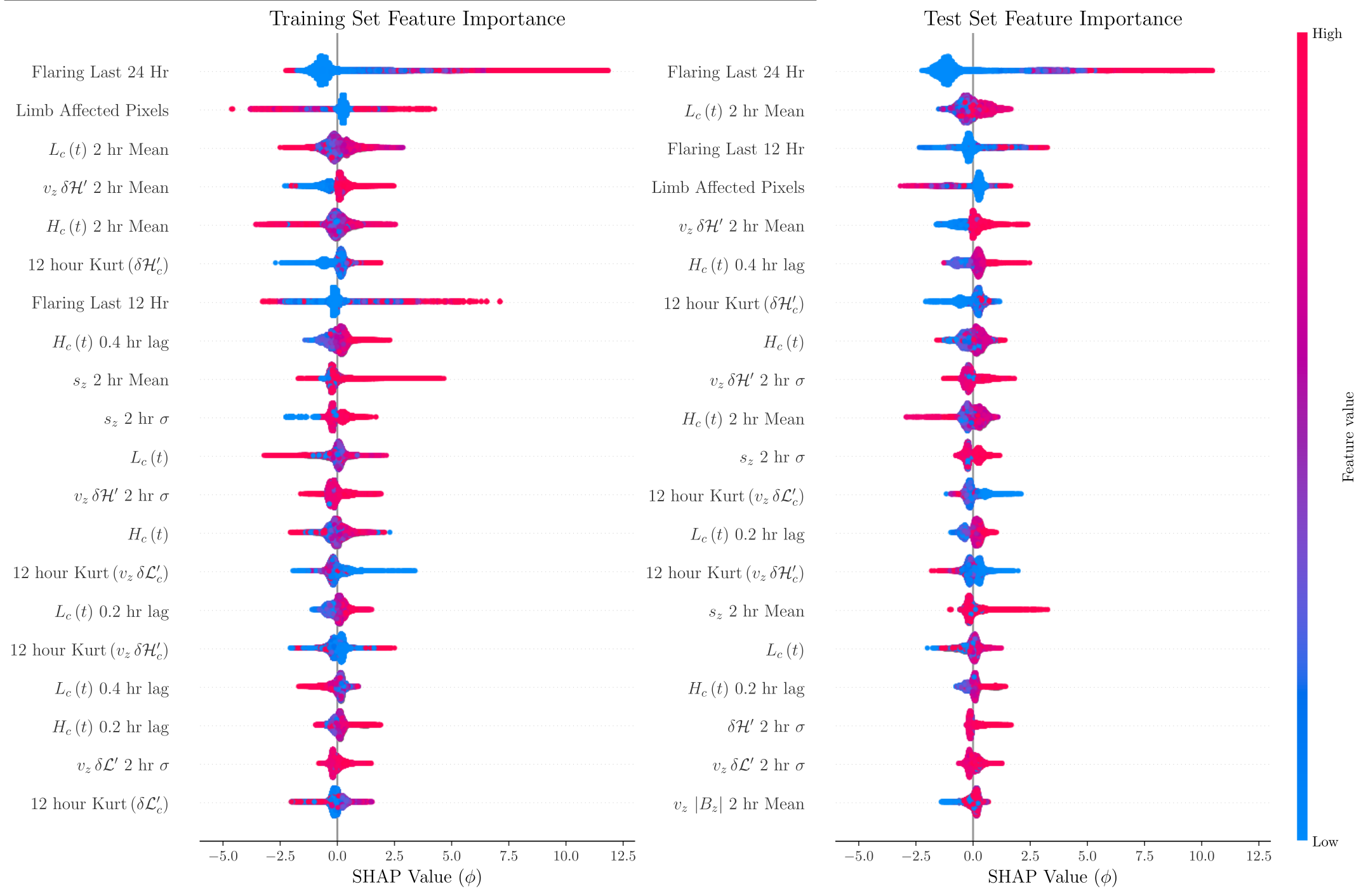}}
\caption{SHapley Additive exPlanations (SHAP) analysis \citep{lundberg2017unified} summary plots of the \texttt{XGBoost} classification model for the Training (left) and Validation (right) Sets. The top 20 most important features are shown in descending order where the feature value increases from low (blue) to high (red). This indicates how each value of the features positively/negatively impacts the magnitude of the prediction by increasing/decreasing the SHAP Value, $\phi$.}
\label{fig:shapAnal}
\end{figure*}

Upon completion of the ten cross-validation folds for the hyperparameter tuning and final model training, the model performance is evaluated for the validation set. The TSS score of 0.804 obtained for the operational forecasts indicates the model has developed good skill at accounting for class imbalance. Here, the model's hit rate exceeds the false alarm rate by 80.4\,\% with detections predominantly being true positives. The TSS score also indicates that alongside strong discrimination of events, the model is able to meaningfully extract signal from minority cases (in this case, flares), and that the detection threshold has been well optimised. This is further evidenced when we analyse the full 720\,s cadence data (Model Outlook). From this, accurate outcomes are captured by the model for 92.2\,\% of the 21,272 observations that form the validation set.

A further breakdown of the classifier events is given in Figure\,\ref{fig:confusion} for the confusion matrix of the operational forecast for the validation set. When the Model Outlook is sampled to a 24\,hr operational forecast, i.e. the cadence that would be used for a deployed live predictor, the accuracy decreases from 92.2\,\% but remains high at 87.6\,\% likely due to a FP rate of 11.3\%. These FP outcomes may be skewed slightly due to the decreased granularity of the dataset when deployed as operational forecasts, which is explored further in \S\,\ref{sec:results}.

In Figure\,\ref{fig:shapAnal}, SHapley Additive exPlanations (SHAP) analysis \citep{lundberg2017unified} summary plots are presented for the 20 most important features in the model results for the training (left) and validation (right) sets. The sign of the SHAP value ($\phi$) indicates whether a positive-outcome prediction, i.e. a flare, is more (positive $\phi$) or less (negative $\phi$) likely. The feature values increase from low (blue) to high (red), which along with their respective position on the x-axis, indicates to what degree each feature impacts $\phi$. In the two datasets there are relatively consistent feature rankings, which implies that there is good generalisation of the model. As expected, there is some deviance between the feature ordering and pattern in the training and validation sets. This indicates that firstly, the model has been appropriately tuned to prioritise physically plausible features for flaring. Secondly, these results suggest that the model has a good chance to be robust on unseen data across different periods of the solar cycle.

It is evident that the most prominent feature in the model is the flaring history, with the previous 24 hour history taking top spot in both datasets, whilst also providing the largest range in $\phi$. In both sets, large clusterings of low values are seen to negatively impact $\phi$, indicating that previously non-flaring regions have a strong weighting to remain non-flaring in the predictions made by the model. Typical values of the accumulated winding and helicity ($\mathcal{L}_c$ and $\mathcal{H}_c$) over the previous 2 hours have minimal impact to $\phi$ with a large cluster near 0. However, when $\mathcal{L}_c$ and $\mathcal{H}_c$ are large, they can also lead to large positive and negative $\phi$ values in the training set, indicating that decision trees are combining these (and other) parameters to make new metrics. Similarly to WPM, the kurtoses and other time series statistics such as the running mean and variances are capturing evolution patterns within the data that are seemingly meaningful for flare prediction.

The other quantity the authors wish to draw attention to is the limb-affected pixels. As has been previously mentioned, this feature gives the percentage on the number of pixels that exceed longitudes of $\pm60^{\circ}$ relative to the central meridian. In both the training and validation sets, the low values (i.e. when no projection effects are present) only marginally increase $\phi$, whilst large values (i.e. when there are projection effects present) have large impacts on the predicted outcome; for both positive (flaring) and negative (non-flaring) outcomes. This suggests that the \texttt{XGBoost} algorithm is making connections between the behaviour of topological quantities and how much of the SHARP is subjected to projection effects. This will be discussed in more detail later.

\section{Results \& Analysis}\label{sec:results}
\begin{figure*}
\centerline{\includegraphics[width=0.85\textwidth]{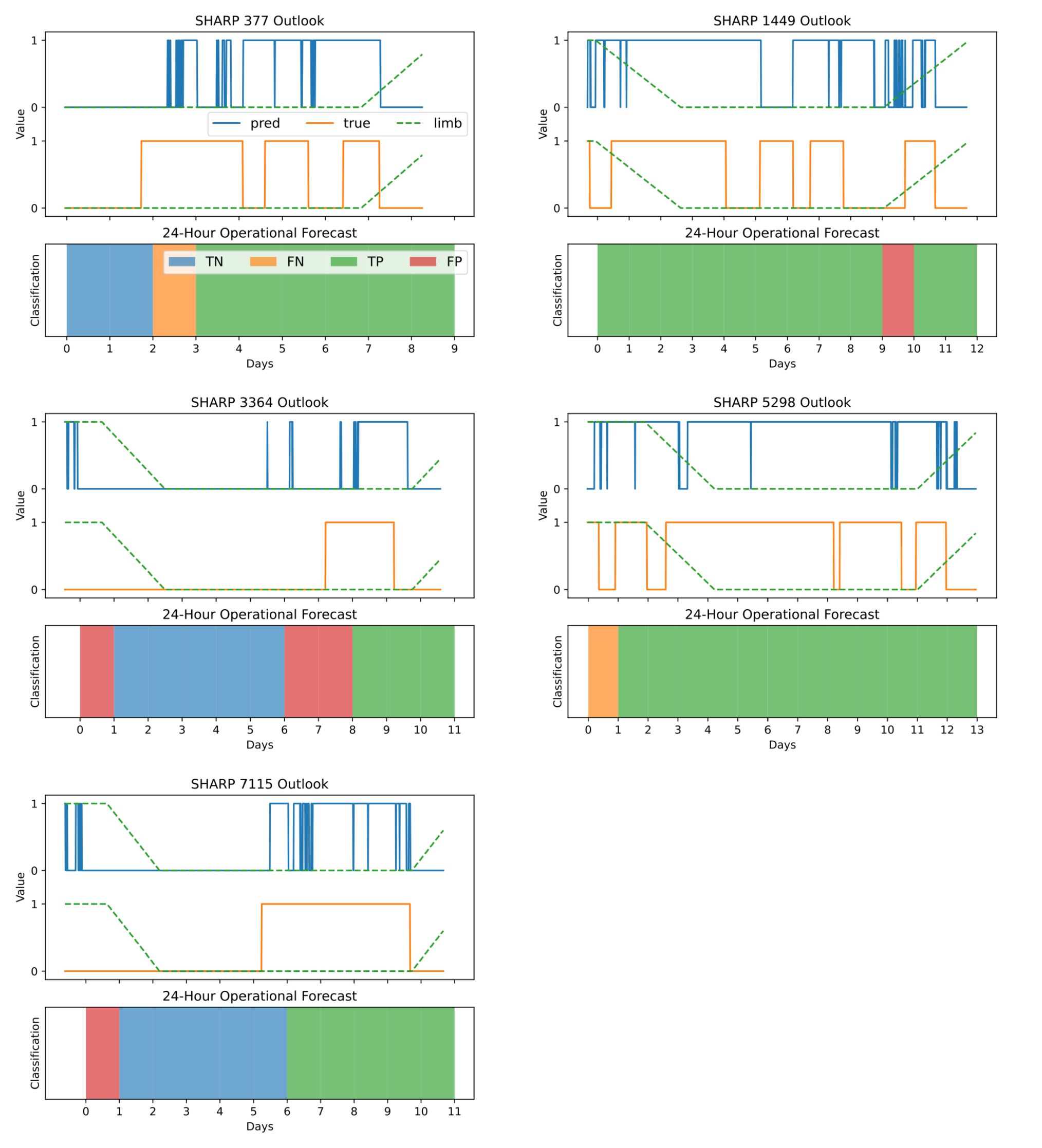}}
\caption{The Model Outlook and operational forecasts are provided for the flaring regions in the validation set. Here, the outlook provides a yes/no outcome for flaring in the following 24\,hours from each SHARP observation. The ground truth is shown in orange, whilst the model prediction is shown in blue. The percentage of limb-affected pixels for the SHARP region is indicated in dashed green. The corresponding operational forecast shown below the Model Outlook collates these 720\,s cadence predictions, and if a positive outcome exists within a 24\,hour period (midnight-to-midnight), determines that a flare is likely the next day. Here, true negative, false negative, true positive, false positive are indicated in blue, red, green, and orange, respectively.}
\label{fig:XCres}
\end{figure*}
\begin{figure*}
\centerline{\includegraphics[width=0.8\textwidth]{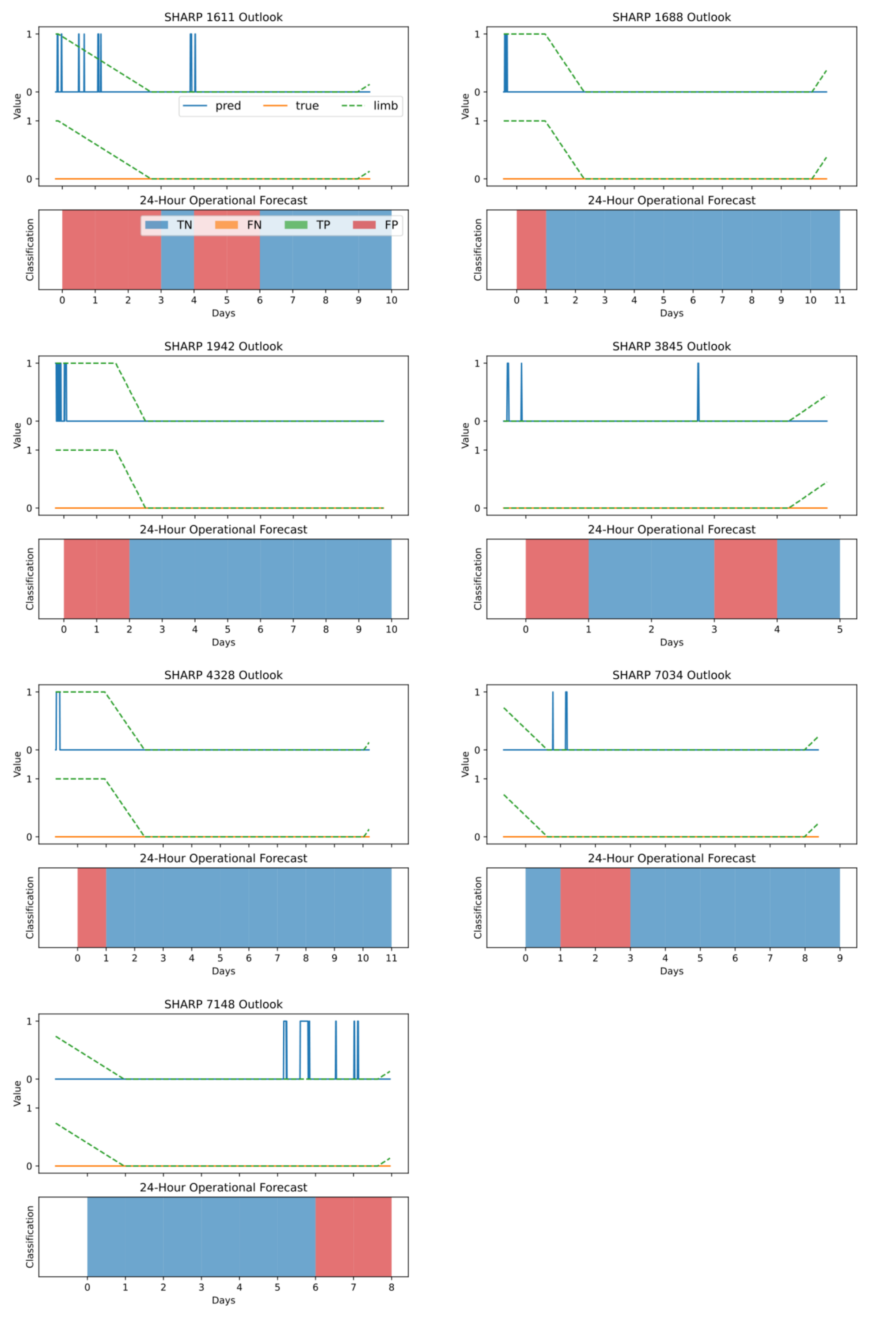}}
\caption{Same as Figure\,\ref{fig:XCres} but for the non-flaring region in the validation set that have FP alerts.}
\label{fig:FPres}
\end{figure*}
\begin{figure*}
\centerline{\includegraphics[width=0.95\textwidth]{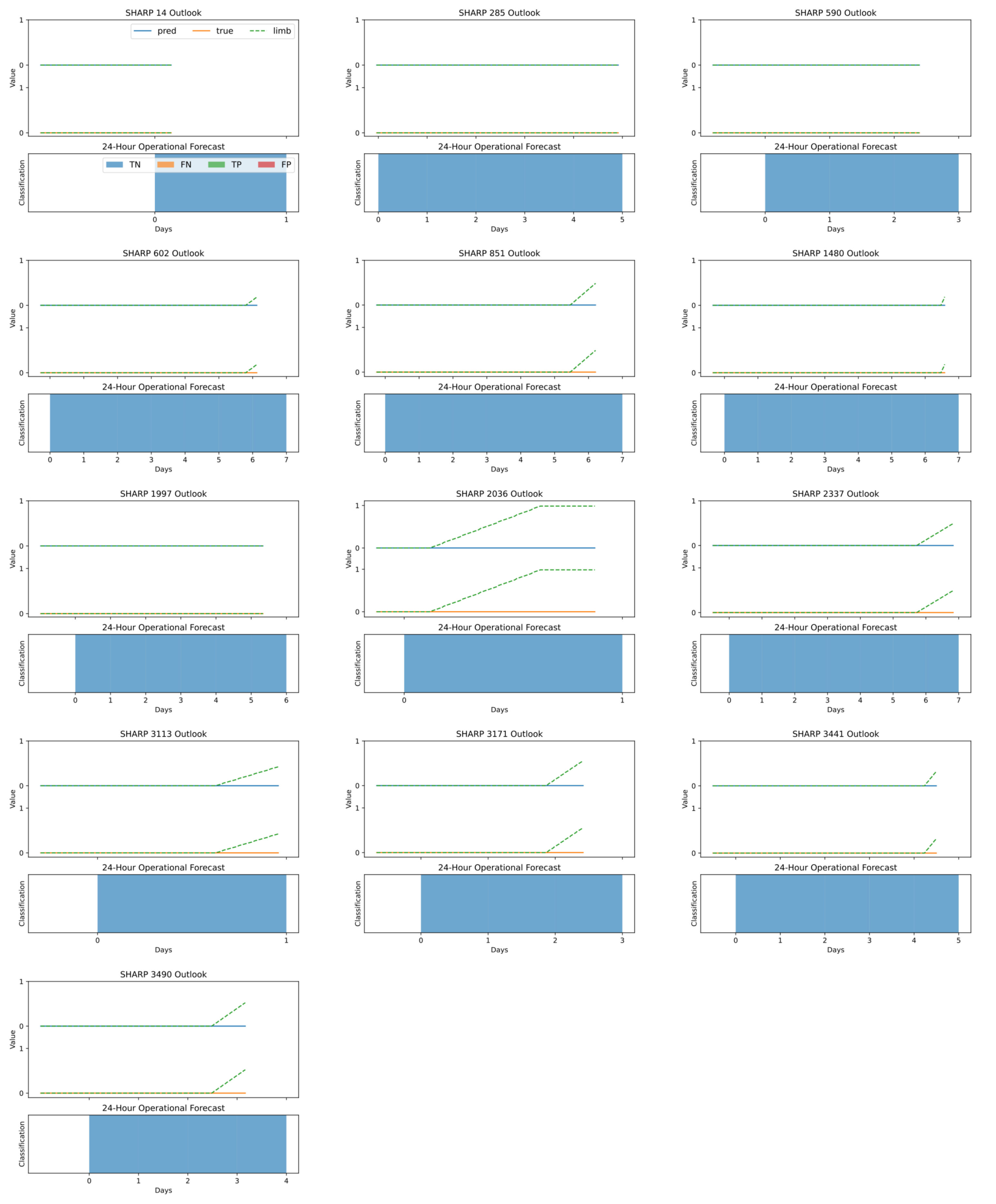}}
\caption{Same as Figure\,\ref{fig:XCres} but for the non-flaring regions in the validation set that have only true negative outcomes.}
\label{fig:NFres}
\end{figure*}
\begin{figure*}
\centerline{\includegraphics[width=\textwidth]{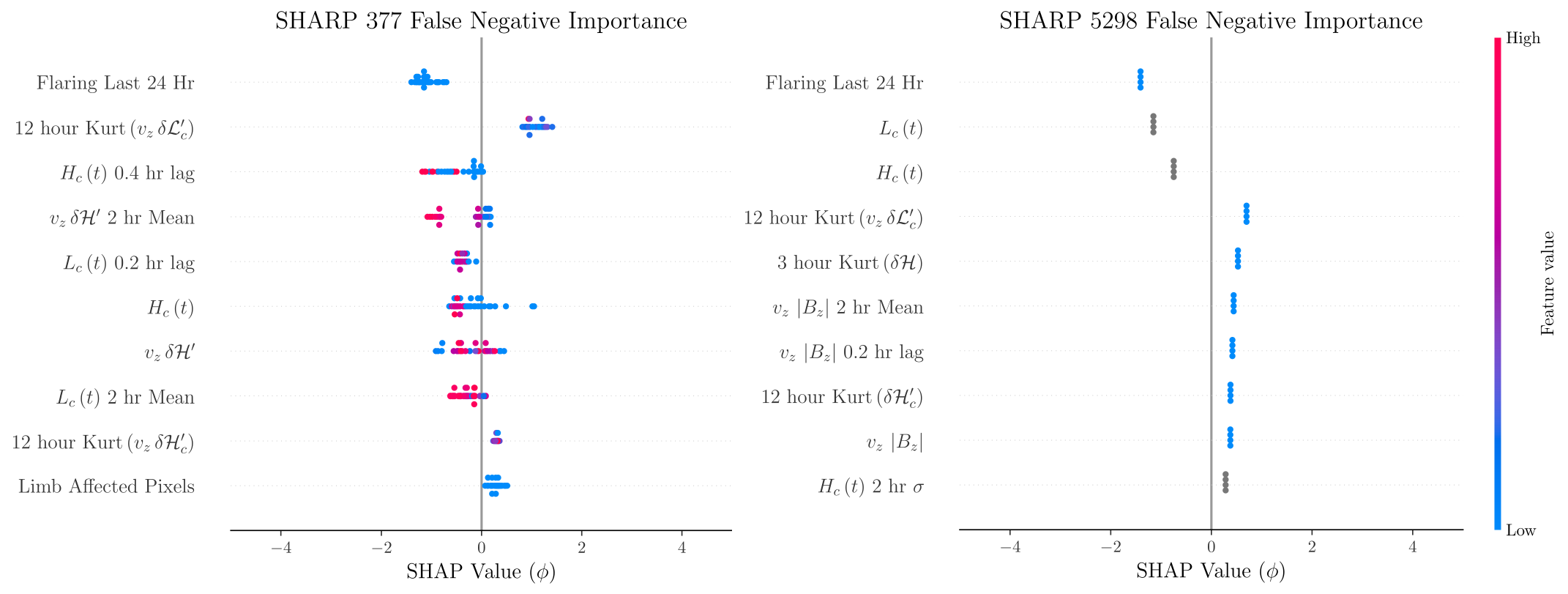}}
\caption{SHAP summary plots for the top 10 features during the periods in the Model Outlook that lead to false negative predictions for SHARP regions 377 and 5298.}
\label{fig:FN}
\end{figure*}
\begin{figure*}
\centerline{\includegraphics[width=\textwidth]{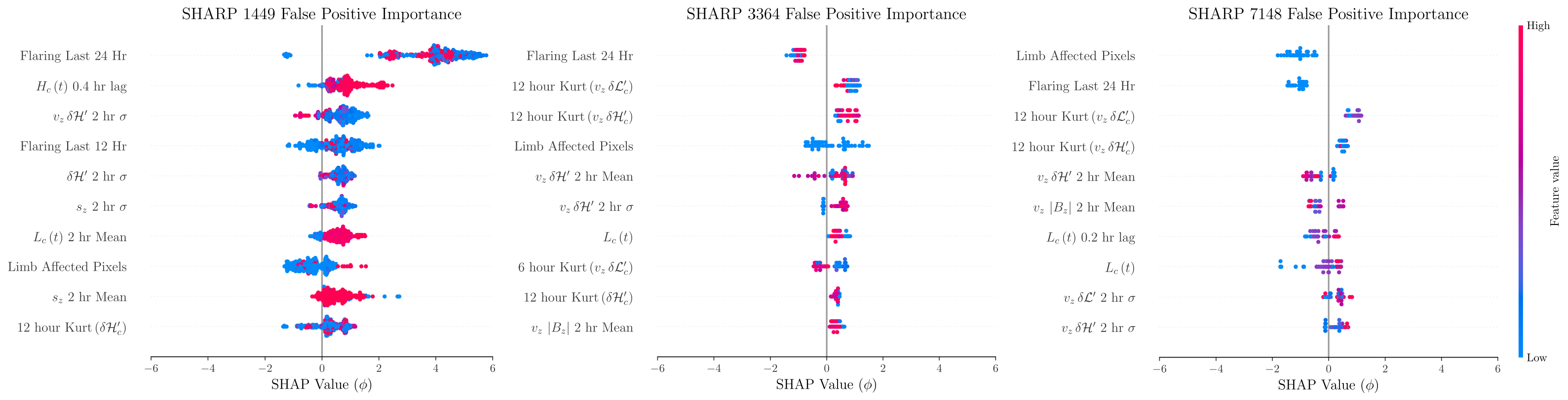}}
\caption{SHAP summary plots for the top 10 features during the periods in the Model Outlook that lead to false positive predictions for SHARP regions 1449, 3364, and 7148.}
\label{fig:FP}
\end{figure*}

In this section, we will first present the results from the validation set before proceeding with the holdout set. In Figure\,\ref{fig:XCres} the flare predictions are presented for 5 flaring regions, whilst the non-flaring regions are shown in Figures\,\ref{fig:FPres} and \ref{fig:NFres}. In these three Figures, each SHARP region is presented across two panels. On the top panel the Model Outlook results are shown, whilst the bottom panel displays the corresponding operational forecast. For the Model Outlook, a yes/no prediction is made every 720\,s about the likelihood of flaring within the next 24\,hours immediately following that SHARP observation time, which is shown in blue. The ground truth is shown in orange and the decimal for the percentage of pixels exceeding longitudes of $\pm60\,^{\circ}$ is shown in dashed green. These outcomes are collated into 24\,hour periods spanning midnight-to-midnight for each day the SHARP region is present. Within each of these 24\,hour periods, if the Model Outlook has a positive prediction for one of the observation periods, the operational forecast determines that a flare is the likely outcome for the following day. How this prediction compares to the known flaring outcome for the SHARP region is then plotted on the right-hand panel for each SHARP. In these plots, TN, FN, TP, and FP outcomes are indicated in blue, red, green, and orange, respectively.

Across the 5 regions shown in Figure\,\ref{fig:XCres}, the operational forecast makes correct predictions (TN + TP) for flaring outcomes the following day for 87.5\,\% of the days. Here, two FN outcomes are made, both at the onset of flaring, for SHARPs 377 and 5298. For SHARP 5298, this occurs in a period where the entirety of the SHARP data are at longitudes beyond $-60\,^{\circ}$. Furthermore, as can be seen in Figure\,\ref{fig:FN}, only 4 data points exist for this initial prediction and so accumulated and running metrics employed by the model have not yet matured for the region. For SHARP 377, Figure\,\ref{fig:FN} shows that the SHAP summary plots for the top 10 metrics indicate that they all mostly negatively impact $\phi$, which lead to negative (non-flaring) predictions. Interestingly, the 12 hour running kurtosis for \vzdl\ all positively impact $\phi$, suggesting that a change in the active region's magnetic winding has been observed that forewarns of large flares.

In a similar vein to SHARP 5298, false positive detections are seen for SHARPs 3364 and 7115 at the start of their observation periods where they are effected by projection effects. For SHARP 7115, the onset of flaring is then accurately captured, something that was shown to be difficult for an anonymous model in a collaborative workshop \citep{park2020comparison}. For SHARP 3364, we can see there are a few anomolous spikes in the Model Outlook prior to the onset of flaring that lead to FP predictions in the operational forecast for the two days preceding M/X-class flaring. The subsequent SHAP analysis for these FP periods (Figure\,\ref{fig:FP}) indicates that -- like with SHARP 1449's FP prediction for day 9 -- the majority of the metrics contribute to a positive $\phi$. Unlike SHARP 377 where the 12\,hour running \kurtvzdl\ correctly indicates to flaring, for SHARP 3364 it is the main contributor to the FP prediction alongside the 12\,hour \kurtvzdh. 

As for the predictions made for the non-flaring regions, these are segregated into regions that contain TN\,+\,FP (Figure\,\ref{fig:FPres}) and those that only contain TN (Figure\,\ref{fig:NFres}) predictions. Overall, the model performs remarkably well, with a prediction accuracy of 87.6\,\%, i.e. 87.6\,\% of the predictions made correspond to either TN or TP outcomes when compared to the known flaring history of the regions analysed.

As with the flaring regions, many of the FP predictions made for the non-flaring regions correspond to periods where projection effects are present, with the most notable exception being SHARP 7148. Once again employing SHAP analysis for these FP periods (Figure\,\ref{fig:FP}), it is evident that the 12\,hour running kurtoses for \vzdl\ and \vzdh, alongside other running and accumulated helicity-/winding-based metrics lead to the FP predictions. This, alongside the insights gleaned from SHAP analysis on SHARPs 1449 and 3364, indicates that perhaps more nuance is required for the interpretation of the changing behaviour of the helicity and winding when it comes to flare prediction. This may be that information on the type of region or the period during the solar cycle the region exists are required alongside the metrics employed in this study. These pieces of information could be incorporated into the \texttt{XGBoost} model, or similarly to the work of \citet{guastavino2022,guastavino2025}, could be implemented in parallel to our \texttt{XGBoost} model, where a top-level algorithm makes a prediction based upon all the data available from the parallel models.

\subsection{Holdout Set}\label{sec:hold}
\begin{figure*}
\centerline{\includegraphics[width=\textwidth]{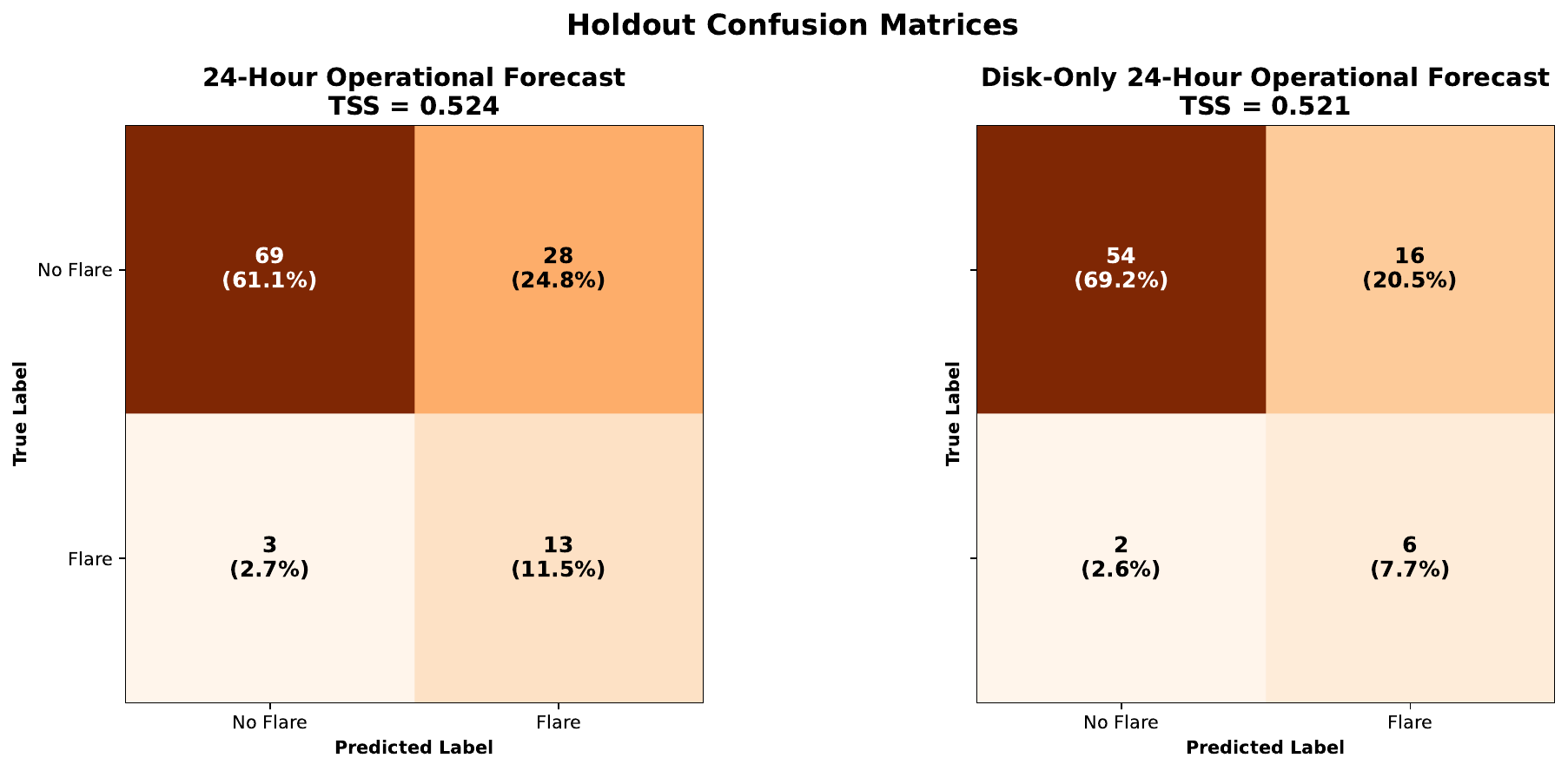}}
\caption{Confusion matrix comparison plots for the \texttt{XGBoost} operational forecasts for all SHARP data in the holdout set. The \textit{left} panel provides results for the full dataset whilst the \textit{right} panel omits limb-affected data. True Skill Statistic scores are shown for illustration.}
\label{fig:confusionHold}
\end{figure*}
\begin{figure*}
\centerline{\includegraphics[width=0.95\textwidth]{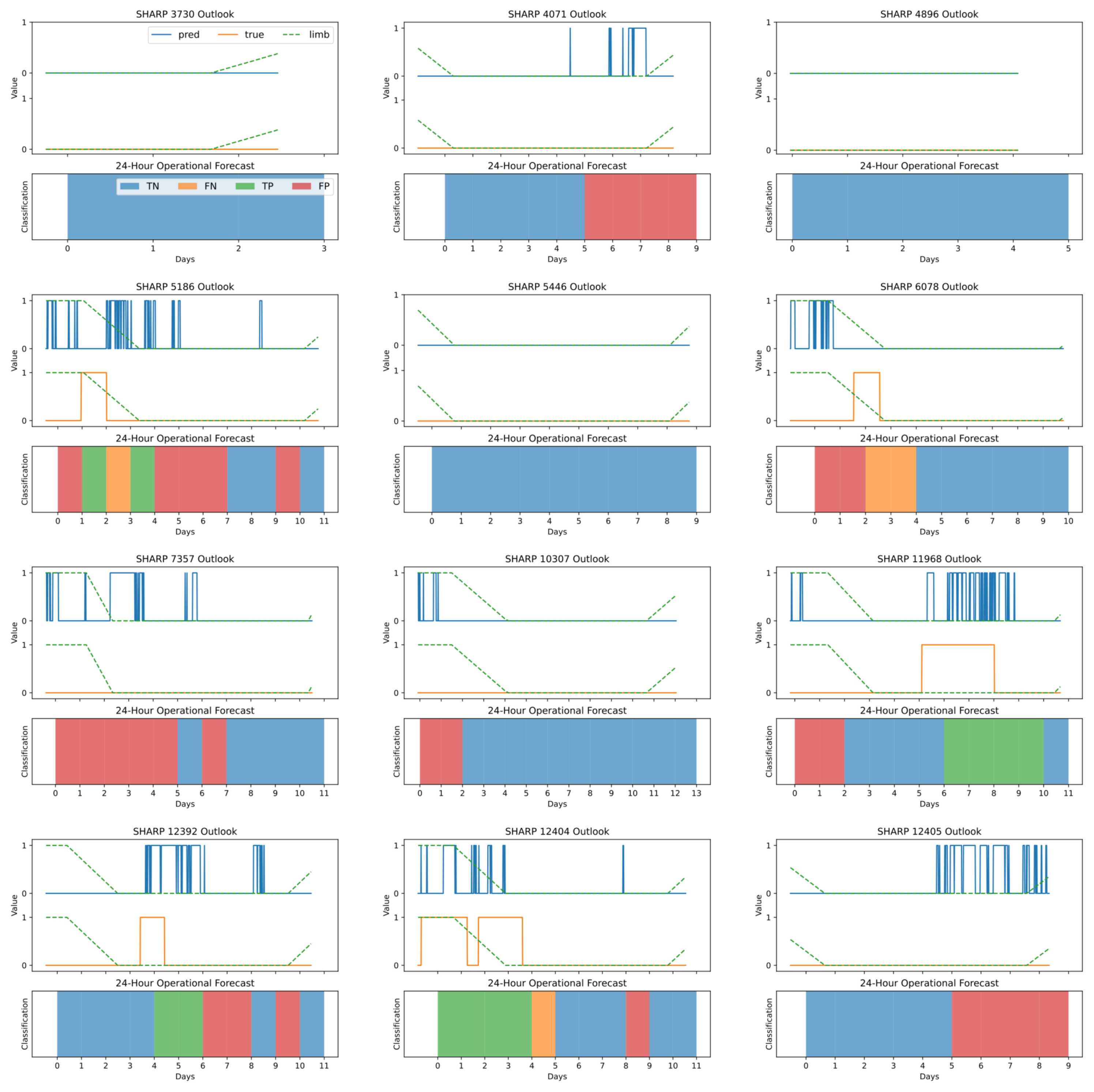}}
\caption{Same as Figure\,\ref{fig:XCres} but for the regions in the holdout set.}
\label{fig:HOres}
\end{figure*}
Following the previous work of this section, the model is retrained to incorporate the validation and training sets so as to be used for deployment. Here, the hyperparameters found during the initial tuning are adopted so as to remain consistent for analysis on a completely unseen holdout set (Table\,\ref{tab:holdoutSet}) of data.\footnote{We note that whilst the validation set is not used to tune hyperparameters nor during the initial training, it is used to define and tune the probability threshold for what constitutes a detection, and so an additional holdout set is required for assessing true unseen data.}

The deployable version of our model is then tested on the holdout set. The confusion matrices for this data are shown in Figure\,\ref{fig:confusionHold} for the operational forecast. Compared to the validation set, we find the TSS score now decreases from 0.804 to \tsssa\ for the operational forecast. For the full dataset (left panel in Figure\,\ref{fig:confusionHold}), the FP rate has increased 13\,\% whilst the TP outcome has decreased 10\,\% for this holdout dataset. These FP periods can be visualised in Figure\,\ref{fig:HOres}. As we see with the validation set, the limb-affected periods are also leading to FP predictions for the operational forecast with the holdout set. If these limb-affected periods are omitted, the fully-trained model yields Model Outlook and operational forecast accuracies of 87.7\,\% and 76.9\,\% compared to 85.2\,\% and 72.6\,\%, respectively for the full data. The TSS scores show no change when the full data (\tsssa) or disk-only data (\TSSalert) are analysed for the operational forecast. However, visual inspection of the Model Outlook and operational forecast predictions in Figure\,\ref{fig:HOres} show that the model performs similarly to the validation set when it is not impacted by projection effects, though there is a greater tendency for FP outcomes due to incorrect short duration positive predictions in the Model Outlook. This could likely be mitigated by employing temporal importance to these short duration predictions, such as defining a criterion where X positive predictions within a certain period are required for them to be counted. Additionally, it is possible that assigning some form of SES on the 24-hour window to weight the importance of more recent predictions more heavily than those nearer the start of the prediction window may also mitigate these occurrences, though this is beyond the scope of this initial study.

Encouragingly, the operational forecasts for SHARPs 11968, 12392, and 12404 (Figure\,\ref{fig:HOres}) accurately capture the start of the flaring phase, and in the case of SHARP 11968, it also accurately determines when flaring desists. For SHARP 12404, a FN outcome exists on the final flaring day, though it is worth noting that this period occurs when the region transitions from longitudes $>60\,^{\circ}$ to $<60\,^{\circ}$. For SHARP 12392, we obtain FP outcomes once flaring has stopped, though it is worth noting that the active region is subjected to numerous C-class flares during this period with C3.7 and C4.0 flares occurring within 0.75\,hr of each other on day 8. Similarly, SHARP 7357 has several larger C-class flares, with the most notable ones occurring on day 2 being C9.9 and C7.3. In combination with other small flares, the flaring history over the previous 12 and 24 hours would indicate that a significant accumulations of topology is being released and that the region may be considered as flaring. As can be seen in the SHAP analysis for the model training (Figure\,\ref{fig:shapAnal}), the previous flare activity is the parameter with the greatest weighting for the model and so the cumulation of these smaller flares may be wrongly influencing the model prediction in instances where numerous C-class flares are observed.

According to Table\,\ref{tab:holdoutSet}, SHARPs 4071 and 12405 both emitted two M-class flares and a number of C-class flares. However, in Figure\,\ref{fig:HOres} it is evident that no flaring occurs for either region. This is because these flares did not occur until the active regions rotated beyond the view of SDO, though it is worth mentioning that they were actively emitting several larger C-class flares when visible with SDO. From day 5 onwards, the operational forecast for SHARP 4071 yields only FP outcomes (Figure\,\ref{fig:HOres}), with some of these alerts being the cause of short duration spikes in the Model Outlook particularly around days 4 -- 6 with a longer spike starting on day 6 that persists into day 7. This longer duration prediction may be subjected to projection effects as the limb-affected pixels does become non-zero towards then end of this period. Furthermore, during these FP outcomes, the active region exhibits numerous C-class flares, with one 5.5\,hr period consisting of C5.3, C2.0, C2.6, C1.8, and a C1.1 flare, whilst another day sees a C9.2 flare. Similarly for SHARP 12405, C6.0 and C4.2 flares occur within 20 minutes of one another on day 5, whilst day 6 sees an accumulation of C3.4, C2.4, and C1.9 flares. There is no further activity then until the very end of day 9 when the first M-class flare (M1.6) erupts. As with SHARP 12392, the steady release of magnetic energy through more frequent, smaller flares for these two active regions will have an impact on the previous flaring activity parameters of the model, which could be the cause of these FP predictions. 

Furthermore it is likely that the model requires refinement by training with more regions where the maximum flare magnitudes are M-class flares. The results presented here are a promising first step in building a live predictor based on \texttt{ARTop} calculations, but it is evident that regions with numerous C-class and smaller M-class flares are currently problematic for the deployed model. Similarly, further improvement is also needed when processing `near-limb' data. As is indicated in the SHAP analysis, the limb-affected pixels parameter does appear to have important weighting within the \texttt{XGBoost} model, though given the results seen for the holdout and validation sets, this alone is unable to mitigate the manner in which topological quantities like the winding and helicity are effected. 

\section{Concluding Remarks}\label{sec:conc}
This manuscript presents a new solar flare prediction model based upon the magnetic topology calculations of \texttt{ARTop} that has been trained on \TF\ SHARP regions and assessed on a further \HoldoutNumber\ regions. To ensure that overlapping windows are avoided during flare prediction -- and subsequently the duplication of data introducing bias into those predictions -- the Model Outlook is collated into distinct prediction windows of a UT day. During the training phase of the model using these distinct prediction windows, a favourable TSS score of 0.804 is obtained on the validation set. The subsequent SHAP analysis performed on the training and validation sets highlight that the model has developed good generalisation, indicating that it has learnt meaningful trends in the underlying data and how they relate to flaring that exceed magnitudes of M1.0 in the next \predictionRange\,hours. On the \HoldoutNumber\ regions that form the holdout set, the TSS score (\tsssa) does decrease but the accuracy (71.7\,\%) remains high for the daily operational forecast predictions. This is despite a number of the regions being problematic due to projection effects and frequent C-class flares resulting in the release of stored magnetic energy. 

Currently, direct comparisons are difficult between competing flare prediction models -- indicating the need for standardised data, such as that outlined in \citet{hollanda21}, to be employed across studies. For example, \citet{bobra2014,cinto20} compare model performance metrics across several studies on flare prediction. In these studies it is shown that most models typically have TSS scores ranging between 0.3 -- 0.7 but as is highlighted in \citet{bobra2014} Table\,2, the intervals between predictions, the data used for evaluation and training all differ. When \citet{bobra2014} tune their model for TSS, they obtain  a score of 0.761 when using an operational model (24\,hour prediction intervals). Similarly, whilst \citet{nishizuka2017} find TSS scores above 0.9 for their model, outperforming NICT space weather forecast center (TSS = 0.21), this calculation is only for 29 X-class flares and did not include limb-affected data/statistics. On the other hand, the CNN model of \citet{pandey23} yields a TSS score of 0.54 for flares exceeding M1.0 within the next 24\,hours. Similarly, Deep Flare Net \citep[DeFN]{nishizuka21} yields TSS scores of 0.80 for flares M1.0 and above and 0.63 for flares C1.0 and above. However, their operational assessment in 2019\,--\,2020 saw scores decrease to 0.24(0.48) for M-class flares when a probability threshold of 50(40)\,\% is adopted.

By comparison, the work presented in this manuscript explores data that contains a total of \cflares\ flares, of which there are \xflares\ X-class and \mflares\ M-class flares. When evaluating the training model with the validation set, we obtain a TSS score of0.804 for the operational forecasts. However, when employing the fully trained model on the holdout set, the score decreases \tsssa, respectively. If the limb-affected data are ignored, such as with \citet{nishizuka2017}, then the TSS scores remains similar for our deployable model (\TSSalert), showing similar predictive efficacy to DeFN. It is worth noting, that unlike many of the models discussed here, the operational forecast and Model Outlook have not been optimised on TSS score but on the weighted combination of statistics given in equation\,\ref{eq:weightedCriterion}.

One of the main issues encountered with the fully trained model evaluation is the digital nature of what constitutes a positive prediction. That is, a hard threshold is set where an outcome is only positive (flaring) if the GOES flux $\ge\mathrm{M1.0}$ for a single flare. In SHARP 4071 we witness numerous C-class flares during a 5.5\,hr period where the total magnitude exceeds M1.0, which due to this digital cut-off, result in FP outcomes. Similarly, SHARP 7357 exhibits a C9.9 flare during a period where the operational forecast also yields a FP outcome. One could argue that the two scenarios encountered here, whilst giving the wrong or undesired outcomes, are not incorrect results but merely artifacts due to the deficiencies in how the solar flare forecasting community defines what a meaningful flare is for space weather. 

The other challenge for the model presented here is posed by regions that rotate into view. Firstly, these regions have likely undergone a significant portion, if not all, of their flux emergence phase during which \texttt{ARTop} will not have captured the build-up of topology (\Hc\ and \Lc), which as is evidenced in our SHAP analysis, forms an important metric in the model's ability to predict flaring. This could perhaps be mitigated by estimating the topology build-up through extrapolations (e.g., \citealp{jarolim2023probing}). The second, and more important issue of regions rotating into view, are the projection effects. Whilst the limb-affected pixels parameter does appear to have important weighting within the \texttt{XGBoost} model, this metric alone is unable to mitigate the manner in which topological quantities like the winding and helicity are effected.

Before a topological based live predictor can be used for operational forecasting of solar flares, these issues must be addressed. In future work, we aim to resolve the issues presented in this manuscript by continuing to update the living dataset first created by WPM, including extrapolations to approximate the topology build-up of active regions that rotate into view, alongside incorporation of multi-layered models such as discussed in \citet{guastavino2022,guastavino2025}. In light of the issues around larger/numerous C-class flares, we will also explore expanding the binary classification of flare/no flare to a multi-class classification in a similar manner to that explored in \citet{bringewald25}.

\begin{acknowledgements}
The authors acknowledge support from the US Air Force grant FA8655-23-1-7247. D. M. acknowledges support from a Leverhulme Trust grant (RPG-2023-182), a Science and Technologies Facilities Council (STFC) grant (ST/Y001672/1) and a Personal Fellowship from the Royal Society of Edinburgh (ID:4282). C.P acknowledges support from The UK Science and Technology Funding Council under grant number ST/W00108X/1. This research utilises version 5.1.2 of the SunPy open source software package \citep{sunpypaper}.
\end{acknowledgements}

\newpage
\bibliographystyle{aasjournal}
\bibliography{ref} 

\begin{appendix}      
Here, the full list of \TrainingNumber\ SHARP regions used for the training of the model are given in Table\,\ref{tab:sharps_appendix}. A breakdown on the largest flare size, as well as the total number of X-, M-, and C-class flares are given for each region. In total, \FAR\ regions exhibit flaring above M1.0, with a further \NF\ regions being flare free.

\begin{center}
\begin{longtable}{ c c c c c c }
\caption{SHARP regions investigated with flare information provided by \textit{HEK}.}\label{tab:sharps_appendix} \\

\hline \multicolumn{1}{c}{NOAA Active Region} & \multicolumn{1}{c}{SHARP Number} & \multicolumn{1}{c}{Largest Flare} & \multicolumn{1}{c}{\# X--class Flares} & \multicolumn{1}{c}{\# M--class Flares} & \multicolumn{1}{c}{\# C--class Flares} \\ 
\endfirsthead
\hline

11069 & 8 & M1.2 & --- & 1 & 7 \\
11071 & 17 & --- & --- & --- & --- \\
11072 & 26 & --- & --- & --- & --- \\
11076 & 43 & --- & --- & --- & --- \\
11073 & 45 & --- & --- & --- & --- \\
11079 & 49 & M1.0 & --- & 1 & --- \\
11080 & 51 & C1.2 & --- & --- & 1 \\
11081 & 54 & M2.0 & --- & 1 & 5 \\
11086 & 83 & --- & --- & --- & --- \\
11087 & 86 & C3.4 & --- & --- & 5 \\
11096 & 116 & --- & --- & --- & --- \\
11098 & 131 & --- & --- & --- & --- \\
11105 & 156 & C3.3 & --- & --- & 2 \\
11114 & 219 & --- & --- & --- & --- \\
11116 & 221 & --- & --- & --- & --- \\
11121 & 245 & M5.4 & --- & 3 & 18 \\
11130 & 274 & --- & --- & --- & --- \\
11136 & 316 & --- & --- & --- & --- \\
11138 & 318 & C1.3 & --- & --- & 1 \\
11141 & 325 & C1.9 & --- & --- & 1 \\
11143 & 335 & --- & --- & --- & --- \\
11148 & 347 & --- & --- & --- & --- \\
11151 & 354 & --- & --- & --- & --- \\
11155 & 366 & --- & --- & --- & --- \\
11156 & 367 & --- & --- & --- & --- \\
11160 & 384 & M1.3 & --- & 4 & 20 \\
11165 & 394 & M5.3 & --- & 6 & 24 \\
11172 & 421 & C1.6 & --- & --- & 2 \\
11173 & 429 & --- & --- & --- & --- \\
11179 & 436 & --- & --- & --- & --- \\
11176 & 437 & M1.4 & --- & 3 & 13 \\
11177 & 438 & --- & --- & --- & --- \\
11198 & 527 & --- & --- & --- & --- \\
11199 & 540 & C6.5 & --- & --- & 5 \\
11206 & 572 & --- & --- & --- & --- \\
11209 & 589 & --- & --- & --- & --- \\
11212 & 595 & --- & --- & --- & --- \\
11221 & 619 & --- & --- & --- & --- \\
11219 & 622 & C5.9 & --- & --- & 2 \\
11223 & 625 & C1.4 & --- & --- & 2 \\
11242 & 686 & --- & --- & --- & --- \\
11245 & 700 & --- & --- & --- & --- \\
11248 & 705 & --- & --- & --- & --- \\
11258 & 713 & --- & --- & --- & --- \\
11261 & 750 & M1.2 & --- & 2 & 26 \\
11273 & 799 & --- & --- & --- & --- \\
11281 & 824 & --- & --- & --- & --- \\
11283 & 833 & X2.1 & 2 & 5 & 13 \\
11300 & 875 & --- & --- & --- & --- \\
0 & 877 & C9.6 & --- & --- & 15 \\
11302 & 892 & X1.9 & 2 & 15 & 31 \\
11311 & 926 & --- & --- & --- & --- \\
11314 & 940 & M1.3 & --- & 2 & 32 \\
11318 & 956 & --- & --- & --- & --- \\
11327 & 982 & --- & --- & --- & --- \\
11326 & 990 & --- & --- & --- & --- \\
11339 & 1028 & X1.9 & 1 & 11 & 47 \\
11341 & 1041 & M1.1 & --- & 1 & 3 \\
11357 & 1080 & C1.8 & --- & --- & 3 \\
11373 & 1170 & --- & --- & --- & --- \\
11380 & 1209 & M4.0 & --- & 3 & 11 \\
11385 & 1232 & --- & --- & --- & --- \\
11398 & 1303 & --- & --- & --- & --- \\
11397 & 1312 & --- & --- & --- & --- \\
11416 & 1389 & C1.5 & --- & --- & 1 \\
11420 & 1399 & --- & --- & --- & --- \\
11434 & 1464 & --- & --- & --- & --- \\
11446 & 1497 & --- & --- & --- & --- \\
11450 & 1528 & C3.1 & --- & --- & 5 \\
11463 & 1558 & C8.9 & --- & --- & 5 \\
11456 & 1564 & --- & --- & --- & --- \\
11460 & 1578 & C3.7 & --- & --- & 5 \\
11464 & 1594 & --- & --- & --- & --- \\
11466 & 1603 & M1.0 & --- & 1 & 4 \\
11476 & 1638 & M5.7 & --- & 11 & 84 \\
11477 & 1644 & --- & --- & --- & --- \\
11523 & 1863 & --- & --- & --- & --- \\
11527/28 & 1877 & C5.0 & --- & --- & 7 \\
11531 & 1886 & C1.7 & --- & --- & 2 \\
11547 & 1943 & --- & --- & --- & --- \\
11548 & 1946 & M5.5 & --- & 5 & 11 \\
11549 & 1948 & --- & --- & --- & --- \\
11554 & 1962 & C7.6 & --- & --- & 5 \\
11562 & 1990 & C8.4 & --- & --- & 2 \\
11560 & 1993 & M1.6 & --- & 1 & 17 \\
11565 & 2007 & C2.3 & --- & --- & 2 \\
11568 & 2017 & C1.7 & --- & --- & 1 \\
11591 & 2121 & --- & --- & --- & --- \\
11598 & 2137 & X1.8 & 1 & 3 & 23 \\
11601 & 2158 & --- & --- & --- & --- \\
11613 & 2191 & M6.0 & --- & 5 & 15 \\
11616 & 2203 & C4.3 & --- & --- & 3 \\
11628 & 2262 & C5.7 & --- & --- & 3 \\
11630 & 2270 & C5.5 & --- & --- & 2 \\
11631 & 2291 & C1.4 & --- & --- & 6 \\
11651 & 2348 & --- & --- & --- & --- \\
11664 & 2425 & --- & --- & --- & --- \\
11668 & 2436 & --- & --- & --- & --- \\
11670 & 2460 & C1.5 & --- & --- & 1 \\
11682 & 2501 & --- & --- & --- & --- \\
11680 & 2504 & --- & --- & --- & --- \\
11696 & 2560 & C2.2 & --- & --- & 2 \\
11697 & 2569 & --- & --- & --- & --- \\
11699 & 2573 & --- & --- & --- & --- \\
11709 & 2594 & --- & --- & --- & --- \\
11706 & 2595 & --- & --- & --- & --- \\
11707 & 2598 & --- & --- & --- & --- \\
11719 & 2635 & M6.5 & --- & 2 & 13 \\
11737 & 2711 & --- & --- & --- & --- \\
11739 & 2716 & M5.7 & --- & 2 & 18 \\
11748 & 2748 & X3.2 & 3 & 3 & 18 \\
11752 & 2754 & C1.3 & --- & --- & 3 \\
11764 & 2822 & --- & --- & --- & --- \\
11768 & 2832 & --- & --- & --- & --- \\
11776 & 2875 & C3.5 & --- & --- & 3 \\
11784 & 2922 & C1.1 & --- & --- & 1 \\
11789 & 2942 & --- & --- & --- & --- \\
11796 & 2976 & --- & --- & --- & --- \\
11807 & 3019 & --- & --- & --- & --- \\
11809 & 3028 & C4.9 & --- & --- & 9 \\
11821 & 3079 & --- & --- & --- & --- \\
11824 & 3097 & --- & --- & --- & --- \\
11831 & 3115 & --- & --- & --- & --- \\
11835 & 3122 & --- & --- & --- & --- \\
11875 & 3291 & X2.3 & 2 & 12 & 57 \\
11874 & 3293 & --- & --- & --- & --- \\
11878 & 3298 & --- & --- & --- & --- \\
11881 & 3309 & --- & --- & --- & --- \\
11882 & 3311 & X2.1 & 2 & 11 & 9 \\
11886 & 3326 & --- & --- & --- & --- \\
11891 & 3344 & M2.3 & --- & 1 & 4 \\
11892 & 3353 & --- & --- & --- & --- \\
11902 & 3386 & --- & --- & --- & --- \\
11905 & 3420 & C3.3 & --- & --- & 6 \\
11911 & 3446 & --- & --- & --- & --- \\
11916 & 3448 & C3.3 & --- & --- & 6 \\
11930 & 3515 & C1.8 & --- & --- & 3 \\
11936 & 3535 & M9.9 & --- & 4 & 31 \\
11942 & 3560 & --- & --- & --- & --- \\
11945 & 3569 & --- & --- & --- & --- \\
11946 & 3580 & M1.0 & --- & 1 & 3 \\
11951 & 3612 & --- & --- & --- & --- \\
11961 & 3635 & --- & --- & --- & --- \\
11962 & 3658 & --- & --- & --- & --- \\
11967 & 3686 & M6.6 & --- & 23 & 66 \\
11978 & 3741 & --- & --- & --- & --- \\
11988 & 3785 & --- & --- & --- & --- \\
11992 & 3806 & --- & --- & --- & --- \\
11996 & 3813 & M9.3 & --- & 5 & 14 \\
11999 & 3821 & --- & --- & --- & --- \\
12009 & 3853 & --- & --- & --- & --- \\
12011 & 3877 & M1.1 & --- & 1 & --- \\
12017 & 3894 & X1.0 & 1 & 3 & 17 \\
12024 & 3907 & C2.4 & --- & --- & 1 \\
12036 & 3999 & M7.3 & --- & 1 & 27 \\
12039 & 4022 & --- & --- & --- & --- \\
12041 & 4023 & --- & --- & --- & --- \\
12048 & 4065 & C3.4 & --- & --- & 1 \\
12050 & 4075 & C1.1 & --- & --- & 1 \\
12066 & 4131 & --- & --- & --- & --- \\
12063 & 4133 & C3.2 & --- & --- & 8 \\
12086 & 4223 & --- & --- & --- & --- \\
12091 & 4224 & --- & --- & --- & --- \\
12089 & 4231 & M1.1 & --- & 1 & 10 \\
12098 & 4284 & --- & --- & --- & --- \\
12105 & 4302 & --- & --- & --- & --- \\
12118 & 4374 & --- & --- & --- & --- \\
12120 & 4376 & --- & --- & --- & --- \\
12192 & 4698 & X3.1 & 6 & 32 & 72 \\
12193 & 4711 & C5.3 & --- & --- & 1 \\
12198 & 4748 & --- & --- & --- & --- \\
12203 & 4764 & C1.7 & --- & --- & 6 \\
12219 & 4868 & C6.5 & --- & --- & 7 \\
12241 & 4941 & M6.9 & --- & 5 & 20 \\
12257 & 5026 & M5.6 & --- & 3 & 23 \\
12268 & 5107 & M2.1 & --- & 6 & 31 \\
12339 & 5541 & X2.7 & 1 & 3 & 53 \\
12371 & 5692 & M7.9 & --- & 7 & 37 \\
12403 & 5885 & M5.6 & --- & 12 & 82 \\
12422 & 5983 & M7.6 & --- & 18 & 61 \\
12529 & 6483 & M6.7 & --- & 1 & 23 \\
12565 & 6670 & M7.6 & --- & 7 & 47 \\
12629 & 6930 & --- & --- & --- & --- \\
12630 & 6937 & --- & --- & --- & --- \\
12634 & 6946 & --- & --- & --- & --- \\
12638 & 6952 & C4.1 & --- & --- & 3 \\
12644 & 6972 & M5.8 & --- & 7 & 20 \\
12659 & 7022 & C3.3 & --- & --- & 1 \\
12662 & 7045 & --- & --- & --- & --- \\
12665 & 7075 & M2.4 & --- & 2 & 27 \\
12671 & 7107 & C7.0 & --- & --- & 14 \\
12672 & 7110 & C7.7 & --- & --- & 5 \\
12680 & 7131 & C3.0 & --- & --- & 2 \\
12887 & 7798 & X1.0 & 1 & --- & 13 \\
13004 & 8195 & M5.7 & --- & 2 & 2 \\
13575 & 10769 & X3.3 & 1 & 5 & 13 \\
13590 & 10856 & X6.3 & 3 & 10 & 58 \\
13644 & 11098 & --- & --- & --- & --- \\
13665 & 11159 & --- & --- & --- & --- \\
13693 & 11291 & --- & --- & --- & --- \\
13699 & 11309 & --- & --- & --- & --- \\
13704 & 11316 & --- & --- & --- & --- \\
13703 & 11317 & M1.0 & --- & 1 & 1 \\
13706 & 11319 & --- & --- & --- & --- \\
13719 & 11420 & M5.7 & --- & 2 & 4 \\
13881 & 12186 & C8.4 & --- & --- & 4 \\
13983 & 12708 & C3.6 & --- & --- & 1 \\
\hline

\end{longtable}
\end{center}

\end{appendix}

\end{document}